\documentclass[a4paper,twoside,prd,showpacs,preprintnumbers,twocolumn,superscriptaddress,nofootinbib]{revtex4-1}

\usepackage{amssymb,amsmath,bm,natbib}
\usepackage{color}
\usepackage{slashed}
\usepackage{graphics}
\usepackage{graphicx}
\usepackage[utf8]{inputenc}
\usepackage[caption=false]{subfig}
\usepackage{hyperref}
\usepackage{url}
\usepackage{dsfont}
\usepackage{float} 
\usepackage{cancel}
\usepackage{xcolor,colortbl}
\usepackage{rotating}
\usepackage[export]{adjustbox}
\usepackage{subfig}
\usepackage{cleveref}

\newcommand{\e}{\ensuremath{\mathrm{e}}}
\renewcommand{\i}{\ensuremath{\mathrm{i}}}

\definecolor{Gray}{gray}{0.7}

\begin{document}
\preprint{CERN-TH-2020-213, PSI-PR-20-24, UZ-TH 55/20}

\title{Searching for Lepton Flavour (Universality) Violation and Collider Signals\\ from a Singly-Charged Scalar Singlet}

\author{Andreas Crivellin}
\email{andreas.crivellin@cern.ch}
\affiliation{CERN Theory Division, CH--1211 Geneva 23, Switzerland}
\affiliation{Physik-Institut, Universit\"at Z\"urich, Winterthurerstrasse 190, CH--8057 Z\"urich, Switzerland}
\affiliation{Paul Scherrer Institut, CH--5232 Villigen PSI, Switzerland}

\author{Fiona Kirk}
\email{fiona.kirk@psi.ch}
\affiliation{Physik-Institut, Universit\"at Z\"urich, Winterthurerstrasse 190, CH--8057 Z\"urich, Switzerland}
\affiliation{Paul Scherrer Institut, CH--5232 Villigen PSI, Switzerland}

\author{Claudio Andrea Manzari}
\email{claudioandrea.manzari@physik.uzh.ch}
\affiliation{Physik-Institut, Universit\"at Z\"urich, Winterthurerstrasse 190, CH--8057 Z\"urich, Switzerland}
\affiliation{Paul Scherrer Institut, CH--5232 Villigen PSI, Switzerland}

\author{Luca Panizzi}
\email{luca.panizzi@physics.uu.se}
\affiliation{Department of Physics and Astronomy, Uppsala University, Box 516, SE-751 20 Uppsala, Sweden}

\begin{abstract}
In recent years, evidence for lepton flavour universality violation beyond the Standard Model has been accumulated. In this context, a singly charged $SU(2)_L$ singlet scalar ($\phi^\pm$) is very interesting, as it can only have flavour off-diagonal couplings to neutrinos and charged leptons, therefore necessarily violating lepton flavour (universality). In fact, it gives a (necessarily constructive) tree level effect in
 $\ell\to\ell^\prime\nu\nu$ processes, while contributing to charged lepton flavour violating only at the loop level. Therefore, it can provide a common explanation of the hints for new physics in
 $\tau\to\mu\nu\nu/\tau(\mu)\to e\nu\nu$ and of the Cabibbo Angle Anomaly. Such an explanation predicts ${\rm Br }[\tau\to e\gamma]$ to be of the order of a few times $10^{-11}$ while ${ \rm Br}[\tau\to e\mu\mu]$ can be of the order of $10^{-9}$ for order one couplings and therefore in the reach of forthcoming experiments. Furthermore, we derive a {novel} coupling-independent lower limit on the scalar mass of $\approx 200\,$GeV by recasting LHC slepton searches. In the scenario preferred by low energy precision data, the lower limit is even strengthened to $\approx300\,$GeV, showing the complementary between LHC searches and flavour observables. Furthermore, we point out that this model can be tested by reinterpreting dark matter monophoton searches at future $e^+e^-$ colliders.
\end{abstract}

\maketitle

\section{Introduction}
\label{Introduction}

While the LHC, in its quest for discovering beyond the Standard Model (SM) physics, has not discovered any new particles directly~\cite{Butler:2017afk,Masetti:2018btj}, intriguing indirect hints for the violation of lepton flavour universality (LFU) were accumulated. In particular, global fits to $b\to s\ell^+\ell^-$~\cite{Aaij:2014pli,Aaij:2014ora,Aaij:2015esa,Aaij:2015oid,Khachatryan:2015isa,ATLAS:2017dlm,CMS:2017ivg,Aaij:2017vbb} and $b\to c\tau\nu$~\cite{Lees:2012xj,Lees:2013uzd,Aaij:2015yra,Aaij:2017deq,Aaij:2017uff,Abdesselam:2019dgh} data point convincingly towards new physics (NP) with a significance of $>\!5\,\sigma$~\cite{Capdevila:2017bsm,Altmannshofer:2017yso,DAmico:2017mtc,Ciuchini:2017mik,Hiller:2017bzc,Geng:2017svp,Hurth:2017hxg,Alok:2017sui,Alguero:2019ptt,Aebischer:2019mlg,Ciuchini:2019usw,Ciuchini:2020gvn} and $>\!3\,\sigma$~\cite{Amhis:2016xyh,Murgui:2019czp,Shi:2019gxi,Blanke:2019qrx,Kumbhakar:2019avh}, respectively. In addition, also the long standing tension in the anomalous magnetic moment of the muon~\cite{Bennett:2006fi,Aoyama:2020ynm} and the deficit in first row Cabibbo Kobayashi Maskawa (CKM) unitarity~\cite{Belfatto:2019swo,Grossman:2019bzp,Coutinho:2019aiy,Crivellin:2020lzu,Coutinho:2020xhc,Crivellin:2020ebi,Kirk:2020wdk,Alok:2020jod,Crivellin:2020oup,Shiells:2020fqp,Seng:2020wjq}, known as the Cabibbo Angle Anomaly (CAA), can be interpreted as signs of LFU violation.

Interestingly, not only the CAA can be explained by a constructive NP contribution to the SM $\mu\to e\nu_\mu\bar\nu_e$ amplitude, but also the analogous tau decays $\tau\to \mu\nu_\tau\bar\nu_\mu$ prefer a constructive NP effect at the $2\,\sigma$ level~\cite{Amhis:2019ckw}. Such an effect can be most naturally generated at tree level, as loop effects are strongly constrained by LEP and LHC data. Furthermore, as data require NP to interfere constructively with the SM, there are only four possible NP candidates\footnote{Also a $SU(2)_L$ triplet scalar gives rise to a SM-like amplitude but interferes destructively.}: vectorlike leptons~\cite{Crivellin:2020ebi}, a left-handed vector $SU(2)_L$ triplet~\cite{Capdevila:2020rrl}, a left-handed $Z^\prime$ with flavour violating couplings~\cite{BCKMM}, and a singly charged $SU(2)_L$ singlet scalar. Interestingly, the last option even gives a necessarily constructive effect and, due to Hermiticity of the Lagrangian, automatically violates lepton flavour (universality). Furthermore, as a singly charged scalar cannot couple to quarks and only generates charged lepton flavour violation at the loop level, it is weakly constrained experimentally by other processes and can therefore potentially explain the CAA and the hints for LFU violation in $\tau$ decays. 
This letter is thus dedicated to the study of the phenomenology of the singly charged $SU(2)_L$ singlet scalar in the light of the hints for LFU violation. 

Singly charged scalars have been proposed within the Babu-Zee model~\cite{Zee:1985id,Babu:1988ki} and studied in  Refs.~\cite{Krauss:2002px,Nebot:2007bc,Cai:2014kra,Cheung:2004xm,Ahriche:2014xra,Chen:2014ska,Ahriche:2015loa,Herrero-Garcia:2014hfa,Herrero-Garcia:2017xdu,CentellesChulia:2018gwr,Babu:2019mfe} as part of a larger NP spectrum, mostly with the aim of generating neutrino masses at loop level. Here, we focus on the SM supplemented only by the singly charged scalar (which constitutes a UV complete model) and perform a comprehensive analysis of flavour and collider constraints in the context of the existing hints for LFU violation.


\section{Model and Observables}
\label{Setup}

As motivated in the Introduction, we supplement the SM by a $SU(2)_L \times SU(3)_C$ singlet $\phi^+$ with hypercharge +1. Interestingly, this allows only for Yukawa-type interactions with leptons
\begin{align}
\mathcal{L} = \mathcal{L}_{\rm SM} - \left({\lambda_{ij}}/{2}\, \bar{L}^c_{a,i}\, \varepsilon_{ab}\, L_{b,j} \, \Phi^+ + {\rm h.c.}\right)\,,
\label{Lag}
\end{align}
but not with quarks. Here $L$ is the left-handed $SU(2)_L$ lepton doublet, $c$ stands for charge conjugation, $a$ and $b$ are $SU(2)_L$ indices, $i$ and $j$ are flavour indices and $\varepsilon_{ab}$ is the two-dimensional antisymmetric tensor. Note that without loss of generality, $\lambda_{ij}$ can be chosen to be antisymmetric in flavour space, $\lambda_{ji}=-\lambda_{ij}$, such that $\lambda_{ii}=0$ and our free parameters are $\lambda_{12}$, $\lambda_{13}$, and $\lambda_{23}$. In addition, there can be a coupling to the SM Higgs doublet $\lambda H^\dagger H \phi^+\phi^-$, which contributes to the mass $m_\phi$ but otherwise only has a significant impact on $h\to\gamma \gamma$.

\begin{figure*}[t!]
\subfloat[]{%
	\includegraphics[width=0.45\textwidth]{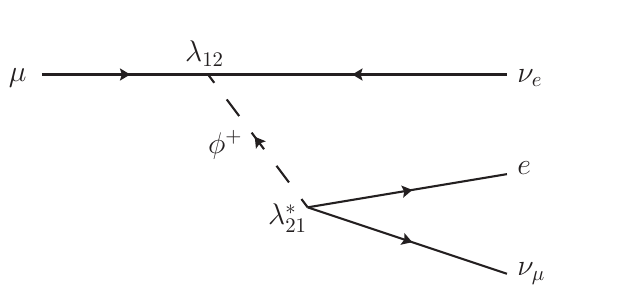}
}
\hfill
\subfloat[]{%
	\includegraphics[width=0.45\textwidth]{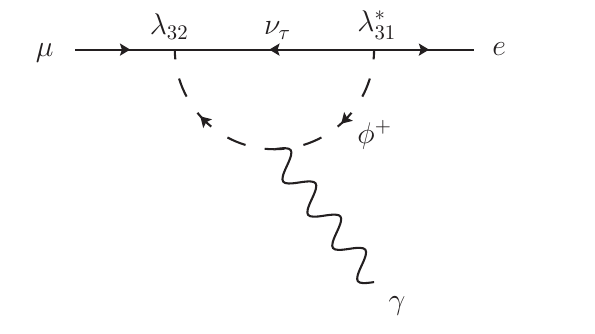}
}\\
\subfloat[]{%
	\includegraphics[width=0.35\textwidth]{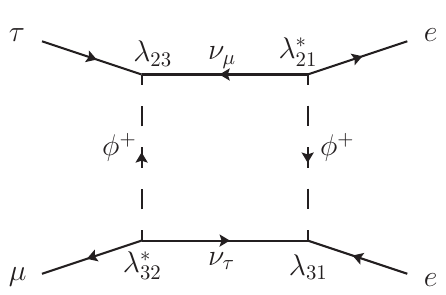}
}
\hfill
\subfloat[]{%
	\includegraphics[width=0.35\textwidth]{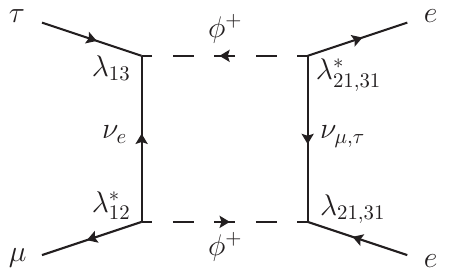}
}
\caption{Feynman diagrams showing the contribution of $\phi^\pm$ to (a) $\mu\to e \nu_\mu \bar\nu_e$, (b) $\mu\to\e\gamma$, and (c, d) $\tau\to \mu ee$. The corresponding diagrams for analogous processes with different flavours are not depicted but can be deduced by straightforward substitutions.\label{FeynmanDiagrams}}
\end{figure*}

\subsection{$\ell\to \ell^\prime\nu\nu$}
The SM decay of a charged lepton into a lighter one and a pair of neutrinos is modified at tree level in our model. Applying Fierz identities (see, e.g., Ref.~\cite{Nieves:2003in}) one can remove the charge conjugation and transform the amplitude to the $V-A$ structure of the corresponding SM amplitude. Taking only into account interfering effects with the SM we have
 \begin{align}
\delta(\ell_i\to \ell_j\nu\nu)=\frac{\mathcal{A}_{NP}(\ell_i\to\ell_j\nu_i\bar\nu_j)}{\mathcal{A}_{SM}(\ell_i\to\ell_j\nu_i\bar\nu_j)} =\frac{\left\vert \lambda_{ij}^2\right\vert}{g_2^2}\frac{m_W^2}{m_\phi^2}\,.
\end{align}
This has to be compared to~\cite{Amhis:2019ckw}
\begin{align}
\begin{split}
{\left. {\frac{{{\cal A}(\tau  \to \mu \nu \bar \nu )}}{{{\cal A}(\mu  \to e\nu \bar \nu )}}} \right|_{{\rm{EXP}}}} &= 1.0029(14){\mkern 1mu} ,\\
{\left. {\frac{{{\cal A}(\tau  \to \mu \nu \bar \nu )}}{{{\cal A}(\tau  \to e\nu \bar \nu )}}} \right|_{{\rm{EXP}}}} &= 1.0018(14){\mkern 1mu} ,\\
{\left. {\frac{{{\cal A}(\tau  \to e\nu \bar \nu )}}{{{\cal A}(\mu  \to e\nu \bar \nu )}}} \right|_{{\rm{EXP}}}} &= 1.0010(14){\mkern 1mu} ,
\end{split}
\end{align}
with the correlations also given in Ref.~\cite{Amhis:2019ckw}.

Furthermore, the effect in $\mathcal{A}(\mu \to e \,\nu_\mu \overline{\nu}_e)$ leads to a modification of the Fermi constant, which enters not only the electroweak (EW) precision observables, but also the determination of $V_{ud}$ from beta decays. Superallowed beta decays provide the most precise determination of $V_{ud}$, leading to~\cite{Seng:2020wjq}\footnote{Alternative determinations can be found in Refs.~\cite{Czarnecki:2019mwq,Shiells:2020fqp}. In addition, there is the possibility of ``new nuclear corrections'' (NNCs)~\cite{Seng:2018qru,Gorchtein:2018fxl}. However, as this issue is debated, we will not consider them here for the sake of argument (i.e., pointing out the potential NP implications). }
\begin{align}
V_{us}^\beta =0.2280(6)\,.
\label{Vusbeta}
\end{align}
This value of $V_{us}^\beta$ can now be compared to $V_{us}$ from kaon~\cite{Aoki:2019cca} and tau decays~\cite{Amhis:2019ckw} 
\begin{align}
\begin{split}
V_{us}^{K_{\mu 3}}&=0.22345(67)\,,\;\;\; V_{us}^{K_{e 3}}=0.22320(61)\,,\\
V_{us}^{K_{\mu 2}}&=0.22534(42)\,,\;\;\;V_{us}^\tau = 0.2195(19)\,,
\end{split}
\label{VusKl3}
\end{align}
which are significantly lower. This is what constitutes the CAA. The tension can be alleviated by the NP effect given by
\begin{align}
V_{us}^\beta  &\equiv \sqrt {1 \!-\! {{(V_{ud}^\beta )}^2} \!-\! |{V_{ub}}{|^2}} \nonumber\\
& \simeq V_{us}^{\rm{L}}\!\left[ {1 + {{\left( {\frac{{V_{ud}^{\rm{L}}}}{{V_{us}^{\rm{L}}}}} \right)}^2}\,\delta (\mu\to e\nu\nu)} \right].
\end{align}
where $V_{us(ud)}^{\rm L}$ is the value appearing in the CKM matrix.
As $G_F$ enters also the calculation of the EW gauge boson masses and $Z$ pole observables, a global fit is necessary. Adding the determinations of the CKM elements to the standard EW observables (see, e.g., Ref.~\cite{Crivellin:2020zul} for details on our input and implementation) calculated by HEPfit~\cite{deBlas:2019okz}, we find
\begin{equation}
\delta (\mu\to e\nu\nu)
=0.00065(15)\,.
\end{equation}

\subsection{$\ell\to \ell^\prime \gamma$}

The singly charged scalar generates $\ell\to\ell^{\prime}\gamma$ (see Fig.~\ref{FeynmanDiagrams}). Using the results of Ref.~\cite{Crivellin:2018qmi} we obtain
\begin{align}
	\mathrm{Br}[\mu \to e \gamma]=\frac{m_\mu^3}{4\pi  \Gamma_\mu}\left(|c_L^{{e}\mu}|^2 +|c_R^{{e}\mu}|^2\right)\,,
\end{align}
with $\Gamma_\mu$ being the total width of the muon, and
\begin{align}
\begin{split}
c_L^{{e}\mu}&=\frac{e\,\lambda_{13}^*\,\lambda_{23}}{384\,\pi^2}\frac{m_e}{m_\phi^2}\,,\;\;
c_R^{{e}\mu}=\frac{e\,\lambda_{13}^*\,\lambda_{23}}{384\,\pi^2}\frac{m_\mu}{m_\phi^2}\,.
\end{split}
\label{on_shell}
\end{align}
In what follows we will neglect the mass of the electron and thus $c_L^{{e}\mu}$.
Similarly, the expressions for $\tau\to \mu (e)\gamma$ can be obtained by a straightforward exchange of indices. The current experimental limits at $90\%$ C.L. are as follows~\cite{Bertl:2006up,Aubert:2009ag,TheMEG:2016wtm}:
\begin{align*}
	\begin{split}
{\rm Br}\!\left[\mu\rightarrow e\gamma\right] &\leq 4.2\times10^{-13}\,,\\
{\rm Br}\!\left[\tau\rightarrow \mu\gamma\right] &\leq 4.4\times10^{-8} \,,\\
{\rm Br}\!\left[\tau\rightarrow e\gamma\right] &\leq 3.3\times10^{-8} \,.
	\end{split}
\end{align*}
Note that, in principle, also contributions to anomalous magnetic moments of charged leptons are generated. However, since the effect in our model is not chiral enhanced, the effect is numerically small and can be safely neglected. Interestingly, note that the $\phi^\pm$ interactions do not generate electric dipole moments (EDMs) (disregarding very small quark and neutrino effects already present in the SM) and therefore automatically agree with the latest very stringent bound on the electron EDM from measurements of Rb atoms~\cite{Andreev:2018ayy}.  

\begin{figure}[t!]
	\includegraphics[width=0.4\textwidth]{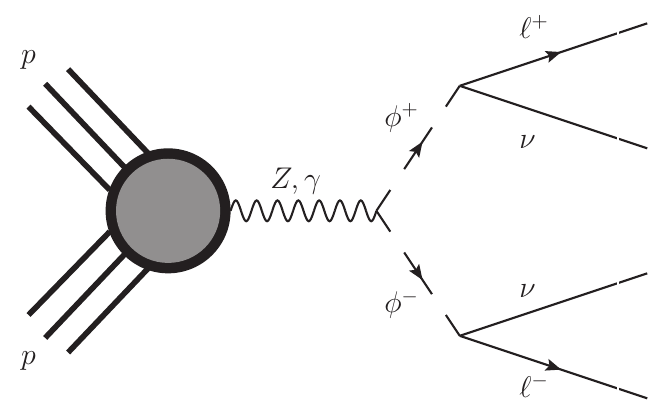} 
	\caption{Diagram showing the Drell-Yan pair production of singly charged scalars. Their decays necessarily give rise to a signal with an oppositely charged lepton pair and missing transverse energy.\label{fig:LHCprocess}}
\end{figure}

\subsection{$\ell\to \ell^\prime\ell^{\prime(\prime)}\ell^{\prime(\prime)}$}
The singly charged scalar contributes to three-body decays to charged leptons at loop level. Here the dominant contribution for sizable couplings $\lambda$ is the box diagram shown in Fig.~\ref{FeynmanDiagrams}. For concreteness, we give the results for $\tau\to3e$ and $\tau\to\mu ee$, while the other decays can be obtained by an appropriate exchange of the flavour indices:
\begin{align}
\mathrm{Br}[\tau \to e \mu \mu] =&\;\frac{m_\tau^5}{1536\,\pi^3 \,\Gamma_{\tau}}
\left \vert\frac{\lambda_{12}^*\,\lambda_{23}\left(\left \vert \lambda_{12}^2\right \vert+\left \vert \lambda_{23}^2\right \vert- \left \vert \lambda_{13}^2\right \vert\right)}{64\,\pi^2 \,m_\phi^2}\right\vert^2\,,\nonumber
\\
\mathrm{Br}[\tau \to eee] =&\;\frac{m_\tau^5}{768\,\pi^3\,\Gamma_{\tau}}\left \vert\frac{\lambda_{12}^*\,\lambda_{23}\left(\left \vert \lambda_{12}^2\right \vert+\left \vert \lambda_{13}^2\right \vert\right)}{64\,\pi^2 \,m_\phi^2}\right\vert^2\,,
\label{tau_decays}
\end{align}
where $\Gamma_\tau$ is the total decay width of the tau. Here we did not include the small on and off shell photon contributions (they are given in Appendix~\cref{app:mu_e_conv}, together with our results for $\mu\to e$ conversion), and we did not give the branching ratios for the decays involving more than one flavour change (such as $\tau\to e\mu e$), which must be tiny in our model due to the measured smallness of $\mu\to e\gamma$. The corresponding experimental bounds (95\% C.L.) are~\cite{Bellgardt:1987du,Hayasaka:2010np,Lees:2010ez,Aaij:2014azz,Amhis:2019ckw} 
\begin{align}
\begin{aligned}
\operatorname{Br}[\mu^- \rightarrow e^- e^+ e^-]&\leq 1.0 \times 10^{-12}\,,\\
\operatorname{Br}[\tau^- \rightarrow e^- e^+ e^-]&\leq 1.4 \times 10^{-8}\,,\\
\operatorname{Br}[\tau^- \rightarrow e^- \mu^+ \mu] &\leq 1.6 \times 10^{-8}\,,\\
\operatorname{Br}[\tau^- \rightarrow \mu^- e^+ e^-] &\leq 1.1 \times 10^{-8}\,,\\
\operatorname{Br}[\tau^- \rightarrow \mu^- \mu^+ \mu^-] &\leq 1.1 \times 10^{-8}\,.
\end{aligned}
\end{align}

\subsection{LHC searches}
\label{LHC}

The singly charged $SU(2)_L$ singlet scalar has the same quantum numbers as the right-handed slepton in supersymmetry. Therefore, bounds from direct searches for smuons and selectrons can be recast to set bounds on our model~\cite{Cao:2017ffm,Alcaide:2017dcx,Alcaide:2019kdr}. The dominant contribution is given by the Drell-Yan pair production of $\phi^\pm$, represented by the Feynman diagram in Fig.~\ref{fig:LHCprocess}. We assume that interference with the SM background (mostly $W^+W^-$ production in this case) can be neglected in the limit of a large enough $m_\phi$ and a narrow $\phi^\pm$ width.

\begin{figure}[t!]
	\includegraphics[width=0.45\textwidth]{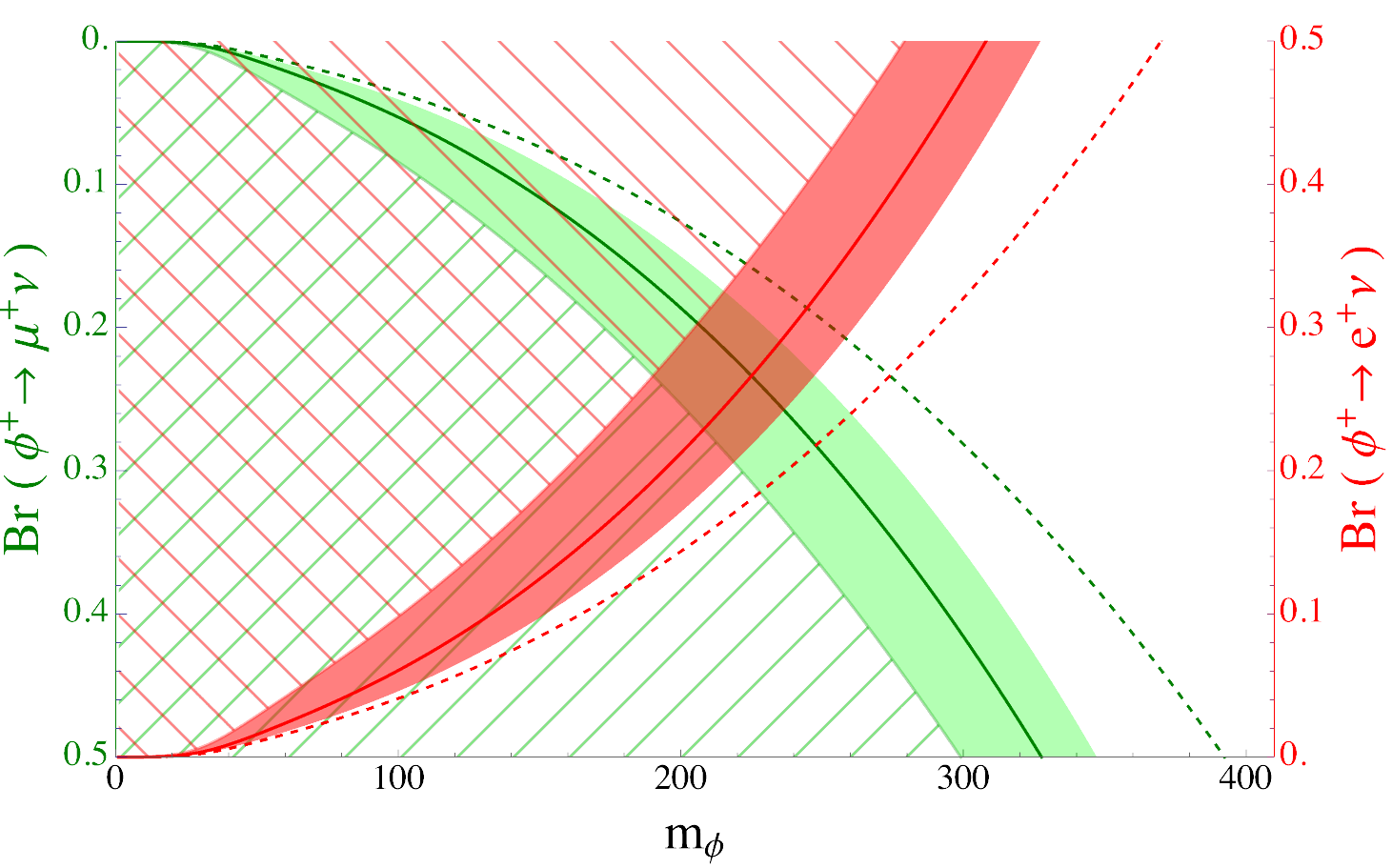} 
	\caption{Recast ATLAS bounds on $m_{\phi}$  and $\rm BR(\phi^+ \to \ell^+\nu)$. The red (green) region is excluded by $e^+e^-$ ($\mu^+\mu^-$) searches (see main text for details). 
		The dashed lines represent the projected exclusion reach for an integrated luminosity of 3~ab$^{-1}$ at the High-Luminosity (HL) LHC. 	\label{fig:LHC_Brs}}
\end{figure}

\begin{figure*}[ht!]
	\includegraphics[width=0.6\textwidth]{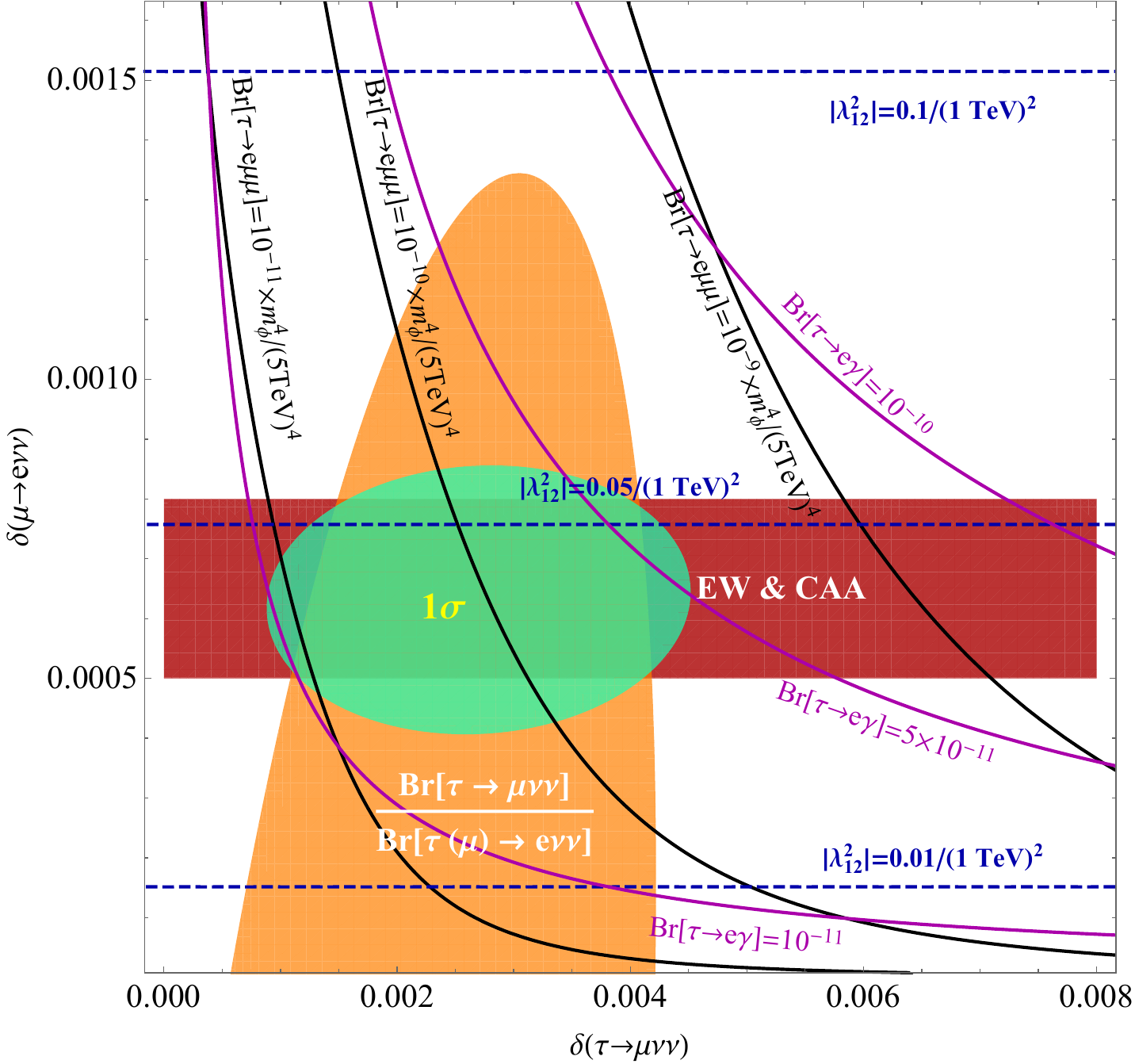} 
	\caption{Preferred regions at the $1\sigma$ level in the $\delta (\tau\to\mu\nu\nu)$--$\delta (\mu \to e\nu\nu)$ plane together with the predictions for $\tau\to e\gamma$ (magenta), $\tau\to e\mu\mu$ (black) and $|\lambda_{12}^2|/m_\phi^2$ (blue) which can be constrained from monophoton searches at future $e^+e^-$ colliders.\label{fig:deltaPlot}}
\end{figure*}

For the reinterpretation of bounds, we consider the most recent ATLAS analysis~\cite{Aad:2019vnb} with 139~fb$^{-1}$ of data, searching for final states with an oppositely charged lepton pair ($e^+e^-$ or $\mu^+\mu^-$) and missing transverse energy. The search targets sleptons decaying into leptons and neutralinos, which corresponds to our setup in the case of a vanishing neutralino mass. The ATLAS bounds on the right-handed slepton mass in this limit is $\approx425$~GeV for both the $e^+e^-$ and $\mu^+\mu^-$ channels and for a 100\% branching ratio of the slepton into the given channel. To reinterpret this result, we have simulated the pair production cross section at leading order with {\sc MG5\_aMC}~\cite{Alwall:2014hca} and rescaled it with a constant $K$-factor, obtained by matching our values with the production cross section to the one given by ATLAS (for a right-handed slepton mass of 500~GeV). A conservative error of $10\%$ has been added on the cross section to account for the differences in the simulation procedures.

Fig.~\ref{fig:LHC_Brs} shows the bounds in the $\rm m_{\phi}$--$\rm Br(\phi^\pm\to e^\pm(\mu^\pm)\nu)$ plane extracted from the analysis of the $e^+e^-$ and $\mu^+\mu^-$ channels of ATLAS. The red (green) hatched region is excluded by the $e^+e^-$ ($\mu^+\mu^-$) channel. The colored bands indicate the change in the limit obtained by linearly varying the efficiency calculated on the value of the ATLAS bound by $\pm 40\%$, between 200~GeV and 425~GeV, for $m_\phi$. The solid line corresponds to the estimated limit without taking into account the additional uncertainties discussed above. As, due to the antisymmetry of the couplings, the sum of the branching ratio to muons and electrons can never be smaller than $1/2$, we can set a coupling-independent limit $\approx 200$~GeV on $m_\phi$.

\subsection{Mono Photon Searches}

LEP-searches for dark matter (DM) with monophoton signatures allow us to set a lower limit on $|\lambda_{12,13}^2|/m_\phi^2$. Using the DELPHI analysis of Refs.~\cite{Abdallah:2003np,Abdallah:2008aa} and Ref.~\cite{Fox:2011fx}, we were able to exploit the kinematic distributions to obtain a bound of $\approx 480$~GeV for zero DM mass on the DM mediator mass for unit coupling strength and vectorial interactions (in the effective theory). Taking into account that we have neutrinos and therefore interference with the SM, this translates into a bound of $\approx 1\,$TeV.

Assuming that $m_\phi$ is sufficiently above the LEP production threshold, as suggested by LHC searches discussed above, we can recast these results. Taking into account that we have a left-handed vector current, we find $(|\lambda_{12,13}^2|)/m_\phi^2\lessapprox1/(175\,{\rm GeV})^2$. This bound would be strengthened for $ \lambda_{12}$ and $\lambda_{13}$ simultaneously nonzero, but further weakened as $m_\phi$ approaches the LEP beam energy. Therefore, it is not yet competitive with flavour bounds but could be significantly improved at future $e^+e^-$ colliders.

\section{Phenomenology}

Let us start our phenomenology by considering the NP effect in $\tau\to \mu\nu\nu$ and $\mu\to e\nu\nu$. The currently preferred regions (at the $1\,\sigma$ level) for $\delta (\tau\to\mu\nu\nu)$ and $\delta (\mu\to e\nu\nu)$ is shown in Fig.~\ref{fig:deltaPlot} as the orange and red regions, respectively, while the combined region is shown in green. Note that for any point within the combined region, $\lambda_{13}$ must be vanishingly small in order not to violate the bounds from $\mu\to e\gamma$ or $\mu\to e$ conversion. Therefore, we can neglect its effect in the following.

This means that in this setup ($\lambda_{13}\!\simeq\! 0$) and we have
${\rm Br}(\phi^+\to\mu^+\nu) =0.5$,
which leads to a bound of $\approx 300$~GeV from the $\mu^+\mu^-$ channel. This bound could be further improved at the HL-LHC~\cite{ApollinariG.:2017ojx} (by around 30\%, as shown in Fig.~\ref{fig:LHC_Brs}, where the ATLAS bounds are rescaled for an integrated luminosity of 3~ab$^{-1}$)
or at the Future Circular hadron Collider (FCC-hh)~\cite{Benedikt:2018csr} where, considering the projections for other scenarios in absence of a dedicated analysis, we estimate a potential improvement of up to a factor of few~\cite{Baumholzer:2019twf}. 
Furthermore, we can correlate $\delta (\tau\to\mu\nu\nu)$ and $\delta (\mu\to e\nu\nu)$ directly to $\tau\to e\gamma$, as indicated by the magenta lines in Fig.~\ref{fig:deltaPlot}, while ${\rm Br}[\tau\to e\gamma]\approx0$. The predicted branching ratio for $\tau\to e\gamma$ is of the order of a few times $10^{-11}$. Furthermore, we can also obtain correlations with $\tau\to 3e$ and $\tau\to e\mu\mu$. Since the branching ratio of the latter is predicted to be larger (for the region preferred by data), we depict it in Fig.~\ref{fig:deltaPlot} as black lines. However, here the correlations are not direct since they depend on $m_\phi$, and we find ${\rm Br}[\tau\to e\mu\mu]\approx 10^{-10}m_\phi^4/(5\,{\rm TeV})^4$. Interestingly, this lies within the reach of BELLE~II~\cite{Inami:2016aba} or the Future Circular electron-positron Collider (FCC-ee)~\cite{Pich:2020qna}. We also depict constant values of $|\lambda_{12}^2|/m_\phi^2$ as dashed blue lines. Even though their values are significantly below the LEP bounds discussed above, future $e^+e^-$ colliders like the International Linear Collider (ILC)~\cite{Baer:2013cma}, the Compact Linear Collider (CLIC)~\cite{Aicheler:2012bya}, the Circular Electron Positron Collider (CEPC)~\cite{An:2018dwb} or the FCC-ee~\cite{Abada:2019zxq} could test the predicted monophoton signature. In particular, the ILC can improve the bound on the Wilson coefficient by a factor of 50~\cite{Habermehl:2020njb}, CEPC by a factor 40~\cite{Liu:2019ogn} and even bigger improvements could be expected at CLIC and at FCC-ee, for which a dedicated study is strongly motivated.

\section{Conclusions}
\label{conclusions}

{The intriguing hints for LFU violation acquired within recent years provide a very promising avenue to search for physics beyond the SM. In this context we studied the phenomenology of the singly charged $SU(2)_L$ singlet scalar which can naturally account for $\tau\to\mu\nu\nu/\tau(\mu)\to e\nu\nu$ and the CAA: the singly charged scalar has only three free couplings (due to Hemiticity of the Lagrangian), necessarily violates lepton flavour and can lead to lepton flavour universality violation if the three couplings are not equal, and leads to a positive definite effect in $\ell\to\ell^\prime\nu\nu$ as preferred by data. Furthermore, the absence of a (pure) NP contribution to the otherwise so stringently constraining electron EDM is guaranteed.} 

Recasting ATLAS searches for right-handed sleptons we derive a {novel} coupling independent limit of $m_\phi\approx200\,$GeV. In the region preferred by LFU violation in tau decays and the CAA, $\lambda_{13}\approx0$ is required by $\mu\to e\gamma$, leading to an LHC bound of $m_\phi\approx300\,$GeV. Concerning lepton flavour violation, we predicted  ${\rm Br}[\tau\to e\gamma]$ to be of order of a few times $10^{-11}$ and ${\rm Br}[\tau\to e\mu\mu]\approx 10^{-10}m_\phi^2/(5\,{\rm TeV})^2$. Therefore, our model can be tested not only by future experiments searching for these LFV decays, but also via direct searches at the High-Luminosity (High-Energy) LHC and FCC-hh and by monophoton searches at future $e^+e^-$ colliders. In particular, the FCC-hh could improve the bound on $m_\phi$ and push the predicted value for ${\rm Br}{[\tau\to e\mu\mu]}$ towards the region observable by BELLE II and FCC-ee, providing a prime example of complementarity between low energy precision experiments and direct searches for NP.

\begin{acknowledgements}
{We thank Joachim Kopp and Marc Montull for useful discussions. The work of A.C., C.A.M. and F.K. is supported by a Professorship Grant (No. PP00P2\_176884) of the Swiss National Science Foundation. A.C. thanks CERN for the support via the scientific associate program. The work of L.P. is supported by the Knut and Alice Wallenberg foundation under the SHIFT project, Grant No. KAW 2017.0100. }
\end{acknowledgements}

\appendix

\section{$\mu\to e$ Conversion}
\label{app:mu_e_conv}
The SM-contributions to $\mu \to e$ conversion can safely be neglected. We parametrise the NP contributions by the effective Lagrangian
\begin{align*}
\mathcal{L}_{\text{eff}}=\sum_{q=u,d}\left(C_{qq}^{V,LL}O_{qq}^{V,LL}+C_{qq}^{V,LR}O_{qq}^{V,LR}\right)+\mathrm{h.c.}\,,
\end{align*}
with
\begin{align}
O_{qq}^{V,LL(R)}&=(\bar{e}\gamma^\mu P_L\mu)(\bar{q}\gamma_\mu P_{L(R)} q),
\end{align}
The singly charged scalar contributes to this process via the off shell photon penguin. In this case the vectorial nature of the photon coupling leads to $C_{qq}^{V,LL}=C_{qq}^{V,LR}$. At leading order we have
\begin{align}
\begin{split}
C_{qq}^{V,LL}&=\frac{e^2 Q_q}{288\,\pi^2 \,m_{\phi}^{2}}\lambda_{13}^*\, \lambda_{23}\,,
\end{split}
\label{off_shell}
\end{align}
where $Q_q$ is the electric charge of the quarks ($Q_u=+\frac{2}{3},\; Q_d=-\frac{1}{3}$).
\\
The transition rate $ \Gamma_{\mu\to e}^N\equiv \Gamma(\mu N\to eN)$ is given by 
\begin{align}
\Gamma_{\mu\to e}^N =&\,4 \,m_\mu^5 \;\Bigg \vert \sum_{q=u,d}\left(C_{qq}^{V,RL}+C_{qq}^{V,RR}\right)\notag\\
&\times\left(f_{Vp}^{(q)}V_N^p\,
+\, f_{Vn}^{(q)}V_N^n\right)
\Bigg\vert^2
+\left(L\leftrightarrow R\right).
\end{align}
The nucleon vector form factors are the same as the ones measured in elastic electron-hadron scattering, i.e.,
\begin{align}
f_{Vp}^{(u)}=2,\;f_{Vn}^{(u)}=1,\;f_{Vp}^{(d)}=1,\;f_{Vn}^{(d)}=2\,.
\end{align}
The overlap integrals $V_{p/n}^N$ depend on the nature of the target $N$. We use the numerical values \cite{Kitano:2002mt}
\begin{align}
\begin{split}
V_{\text{Au}}^p&=0.0974\,,\quad V_{\text{Au}}^n=0.146\,.
\end{split}
\end{align}
The branching ratio of $\mu\to e$ conversion is defined as the transition rate divided by the $\mu$ capture rate:
\begin{align}
\mathrm{Br}(\mu \to e)={\Gamma^\text{conv}}/{\Gamma^\text{capt}}\,,
\end{align}
and for the latter we use \cite{Suzuki:1987jf}
\begin{align}
\Gamma_{\rm Au}^{\rm capt}=8.7\times 10^{-18}\; \text{GeV}\,, \quad \Gamma_{\rm Al}^{\rm capt}=4.6\times 10^{-19}\; \text{GeV}\,.
\end{align}
The experimental limit on $\mu\to e$ conversion is\cite{Bertl:2006up}
\begin{align}
\frac{\Gamma_{\rm Au}^\text{conv}}{\Gamma_{\rm Au}^\text{capt}}&<7.0\times 10^{-13} \quad \text{SINDRUM II}\,.
\end{align}
\\
Adding the on shell (see Eq.~\ref{on_shell}) and off shell (see Eq.~\ref{off_shell}) photon contributions to the $\tau$-decays of Eq.~\ref{tau_decays}, we obtain
\begin{widetext}
	\begin{align}
	{\rm Br}(\tau \to 3e)=&\,
	\frac{e^2 \,m_\tau ^3 }{192\, \pi^3 \,\Gamma_\tau}
	\left| c_R^{e\tau}\right|^2 \left(4 \log \left[\frac{m_\tau^2}{m_e^2}\right]-11\right)\nonumber\\
	&+\frac{m_\tau^5}{3072 \,\pi ^3\, \Gamma_\tau}
	\left(\left|\left(\left| \lambda_{12}\right|^2+\left| \lambda_{13}\right|^2\right)\frac{\lambda_{12}^*\,\lambda_{23}}{32 \,\pi^2 \,m_\phi^2}
	+\,\frac{e^2}{288\,\pi^2}\frac{\lambda_{12}^*\,\lambda_{23}}{m_\phi^2}\right|^2
	+\left|\frac{e^2}{288\,\pi^2}\frac{\lambda_{12}^*\,\lambda_{23}}{m_\phi^2}\right|^2\right)
	\nonumber\\
	&+\frac{e \,m_\tau^4}{384 \,\pi ^3\, \Gamma_\tau} 
	\left(2\left(\left| \lambda_{12}\right|^2+\left| \lambda_{13}\right|^2\right)\frac{ {\rm Re}\left(c_R^{e\tau}\, \lambda_{12} \,\lambda_{23}^*\right)}{32\, \pi ^2\, m_{\phi}^2}
	+\frac{3e^2}{288\,\pi^2}\frac{{\rm Re}\left(c_R^{e\tau} \,\lambda_{12}\, \lambda_{23}^*\right)}{m_\phi^2}\right)\,,
	\\
	{\rm Br}(\tau \to e^\mp \mu^\pm \mu^\mp)=&\,
	\frac{e^2 \,m_\tau^3}{48\, \pi^3\, \Gamma_\tau} \left| c_R^{e\tau}\right|^2 \left(\log \left[\frac{m_\tau^2}{m_\mu^2}\right]-3\right)\nonumber\\
	&+\frac{m_\tau^5}{3072 \,\pi^3 \Gamma_\tau} \left(\left|\left(\left| \lambda_{12}\right|^2-\left| \lambda_{13}\right|^2+\left| \lambda_{23}\right|^2\right)\frac{\lambda_{12}^*\,\lambda_{23}\,}{64\,\pi^2\, m_{\phi}^2} +\frac{e^2}{288\,\pi^2}\frac{\lambda_{12}^*\,\lambda_{23}}{m_\phi^2}\right|^2
	+\left|\frac{e^2}{288\,\pi^2}\frac{\lambda_{12}^*\,\lambda_{23}}{m_\phi^2}\right|^2\right)\nonumber\\
	&+\frac{e\,  m_\tau^4 }{384\, \pi ^5\,  \Gamma_\tau }
	\left(2
	\left(\left| \lambda_{12}\right|^2-\left| \lambda_{13}\right|^2+\left| \lambda_{23}\right|^2\right) \frac{{\rm Re}\left(c_R^{e\tau}\, \lambda_{12}\, \lambda_{23}^*\right)}{64 \,\pi^2 \,m_\phi^2}
	+\frac{3e^2}{288 \,\pi^2}\frac{{\rm Re}\left(c_R^{e\tau} \,\lambda_{12}\, \lambda_{23}^*\right)}{m_\phi^2}\right)\,.\nonumber
	\end{align}
\end{widetext}

\label{tau_decays_photon}
\bibliography{SinglyChargedScalar}

\begin{thebibliography}{99}%
\makeatletter
\providecommand \@ifxundefined [1]{%
 \@ifx{#1\undefined}
}%
\providecommand \@ifnum [1]{%
 \ifnum #1\expandafter \@firstoftwo
 \else \expandafter \@secondoftwo
 \fi
}%
\providecommand \@ifx [1]{%
 \ifx #1\expandafter \@firstoftwo
 \else \expandafter \@secondoftwo
 \fi
}%
\providecommand \natexlab [1]{#1}%
\providecommand \enquote  [1]{``#1''}%
\providecommand \bibnamefont  [1]{#1}%
\providecommand \bibfnamefont [1]{#1}%
\providecommand \citenamefont [1]{#1}%
\providecommand \href@noop [0]{\@secondoftwo}%
\providecommand \href [0]{\begingroup \@sanitize@url \@href}%
\providecommand \@href[1]{\@@startlink{#1}\@@href}%
\providecommand \@@href[1]{\endgroup#1\@@endlink}%
\providecommand \@sanitize@url [0]{\catcode `\\12\catcode `\$12\catcode
  `\&12\catcode `\#12\catcode `\^12\catcode `\_12\catcode `\%12\relax}%
\providecommand \@@startlink[1]{}%
\providecommand \@@endlink[0]{}%
\providecommand \url  [0]{\begingroup\@sanitize@url \@url }%
\providecommand \@url [1]{\endgroup\@href {#1}{\urlprefix }}%
\providecommand \urlprefix  [0]{URL }%
\providecommand \Eprint [0]{\href }%
\providecommand \doibase [0]{http://dx.doi.org/}%
\providecommand \selectlanguage [0]{\@gobble}%
\providecommand \bibinfo  [0]{\@secondoftwo}%
\providecommand \bibfield  [0]{\@secondoftwo}%
\providecommand \translation [1]{[#1]}%
\providecommand \BibitemOpen [0]{}%
\providecommand \bibitemStop [0]{}%
\providecommand \bibitemNoStop [0]{.\EOS\space}%
\providecommand \EOS [0]{\spacefactor3000\relax}%
\providecommand \BibitemShut  [1]{\csname bibitem#1\endcsname}%
\let\auto@bib@innerbib\@empty
\bibitem [{\citenamefont {Butler}(2017)}]{Butler:2017afk}%
  \BibitemOpen
  \bibfield  {author} {\bibinfo {author} {\bibfnamefont {J.~N.}\ \bibnamefont
  {Butler}} (\bibinfo {collaboration} {CMS}),\ }in\ \href@noop {} {\emph
  {\bibinfo {booktitle} {{5th Large Hadron Collider Physics Conference}}}}\
  (\bibinfo {year} {2017})\ \Eprint {http://arxiv.org/abs/1709.03006}
  {arXiv:1709.03006 [hep-ex]} \BibitemShut {NoStop}%
\bibitem [{\citenamefont {Masetti}(2018)}]{Masetti:2018btj}%
  \BibitemOpen
  \bibfield  {author} {\bibinfo {author} {\bibfnamefont {L.}~\bibnamefont
  {Masetti}} (\bibinfo {collaboration} {ATLAS}),\ }\href {\doibase
  10.1016/j.nuclphysbps.2019.03.009} {\bibfield  {journal} {\bibinfo  {journal}
  {Nucl. Part. Phys. Proc.}\ }\textbf {\bibinfo {volume} {303-305}},\ \bibinfo
  {pages} {43} (\bibinfo {year} {2018})}\BibitemShut {NoStop}%
\bibitem [{\citenamefont {Aaij}\ \emph
  {et~al.}(2014{\natexlab{a}})\citenamefont {Aaij} \emph
  {et~al.}}]{Aaij:2014pli}%
  \BibitemOpen
  \bibfield  {author} {\bibinfo {author} {\bibfnamefont {R.}~\bibnamefont
  {Aaij}} \emph {et~al.} (\bibinfo {collaboration} {LHCb}),\ }\href {\doibase
  10.1007/JHEP06(2014)133} {\bibfield  {journal} {\bibinfo  {journal} {JHEP}\
  }\textbf {\bibinfo {volume} {06}},\ \bibinfo {pages} {133} (\bibinfo {year}
  {2014}{\natexlab{a}})},\ \Eprint {http://arxiv.org/abs/1403.8044}
  {arXiv:1403.8044 [hep-ex]} \BibitemShut {NoStop}%
\bibitem [{\citenamefont {Aaij}\ \emph
  {et~al.}(2014{\natexlab{b}})\citenamefont {Aaij} \emph
  {et~al.}}]{Aaij:2014ora}%
  \BibitemOpen
  \bibfield  {author} {\bibinfo {author} {\bibfnamefont {R.}~\bibnamefont
  {Aaij}} \emph {et~al.} (\bibinfo {collaboration} {LHCb}),\ }\href {\doibase
  10.1103/PhysRevLett.113.151601} {\bibfield  {journal} {\bibinfo  {journal}
  {Phys. Rev. Lett.}\ }\textbf {\bibinfo {volume} {113}},\ \bibinfo {pages}
  {151601} (\bibinfo {year} {2014}{\natexlab{b}})},\ \Eprint
  {http://arxiv.org/abs/1406.6482} {arXiv:1406.6482 [hep-ex]} \BibitemShut
  {NoStop}%
\bibitem [{\citenamefont {Aaij}\ \emph
  {et~al.}(2015{\natexlab{a}})\citenamefont {Aaij} \emph
  {et~al.}}]{Aaij:2015esa}%
  \BibitemOpen
  \bibfield  {author} {\bibinfo {author} {\bibfnamefont {R.}~\bibnamefont
  {Aaij}} \emph {et~al.} (\bibinfo {collaboration} {LHCb}),\ }\href {\doibase
  10.1007/JHEP09(2015)179} {\bibfield  {journal} {\bibinfo  {journal} {JHEP}\
  }\textbf {\bibinfo {volume} {09}},\ \bibinfo {pages} {179} (\bibinfo {year}
  {2015}{\natexlab{a}})},\ \Eprint {http://arxiv.org/abs/1506.08777}
  {arXiv:1506.08777 [hep-ex]} \BibitemShut {NoStop}%
\bibitem [{\citenamefont {Aaij}\ \emph {et~al.}(2016)\citenamefont {Aaij} \emph
  {et~al.}}]{Aaij:2015oid}%
  \BibitemOpen
  \bibfield  {author} {\bibinfo {author} {\bibfnamefont {R.}~\bibnamefont
  {Aaij}} \emph {et~al.} (\bibinfo {collaboration} {LHCb}),\ }\href {\doibase
  10.1007/JHEP02(2016)104} {\bibfield  {journal} {\bibinfo  {journal} {JHEP}\
  }\textbf {\bibinfo {volume} {02}},\ \bibinfo {pages} {104} (\bibinfo {year}
  {2016})},\ \Eprint {http://arxiv.org/abs/1512.04442} {arXiv:1512.04442
  [hep-ex]} \BibitemShut {NoStop}%
\bibitem [{\citenamefont {Khachatryan}\ \emph {et~al.}(2016)\citenamefont
  {Khachatryan} \emph {et~al.}}]{Khachatryan:2015isa}%
  \BibitemOpen
  \bibfield  {author} {\bibinfo {author} {\bibfnamefont {V.}~\bibnamefont
  {Khachatryan}} \emph {et~al.} (\bibinfo {collaboration} {CMS}),\ }\href
  {\doibase 10.1016/j.physletb.2015.12.020} {\bibfield  {journal} {\bibinfo
  {journal} {Phys. Lett. B}\ }\textbf {\bibinfo {volume} {753}},\ \bibinfo
  {pages} {424} (\bibinfo {year} {2016})},\ \Eprint
  {http://arxiv.org/abs/1507.08126} {arXiv:1507.08126 [hep-ex]} \BibitemShut
  {NoStop}%
\bibitem [{\citenamefont {ATLAS-CONF-2017-023}()}]{ATLAS:2017dlm}%
  \BibitemOpen
  \bibfield  {author} {\bibinfo {author} {\bibnamefont {ATLAS-CONF-2017-023}}
  (\bibinfo {collaboration} {ATLAS}),\ }\href@noop {} {\ }\BibitemShut
  {NoStop}%
\bibitem [{\citenamefont {CMS-PAS-BPH-15-008}()}]{CMS:2017ivg}%
  \BibitemOpen
  \bibfield  {author} {\bibinfo {author} {\bibnamefont {CMS-PAS-BPH-15-008}}
  (\bibinfo {collaboration} {CMS}),\ }\href@noop {} {\ }\BibitemShut {NoStop}%
\bibitem [{\citenamefont {Aaij}\ \emph {et~al.}(2017)\citenamefont {Aaij} \emph
  {et~al.}}]{Aaij:2017vbb}%
  \BibitemOpen
  \bibfield  {author} {\bibinfo {author} {\bibfnamefont {R.}~\bibnamefont
  {Aaij}} \emph {et~al.} (\bibinfo {collaboration} {LHCb}),\ }\href {\doibase
  10.1007/JHEP08(2017)055} {\bibfield  {journal} {\bibinfo  {journal} {JHEP}\
  }\textbf {\bibinfo {volume} {08}},\ \bibinfo {pages} {055} (\bibinfo {year}
  {2017})},\ \Eprint {http://arxiv.org/abs/1705.05802} {arXiv:1705.05802
  [hep-ex]} \BibitemShut {NoStop}%
\bibitem [{\citenamefont {Lees}\ \emph {et~al.}(2012)\citenamefont {Lees} \emph
  {et~al.}}]{Lees:2012xj}%
  \BibitemOpen
  \bibfield  {author} {\bibinfo {author} {\bibfnamefont {J.}~\bibnamefont
  {Lees}} \emph {et~al.} (\bibinfo {collaboration} {BaBar}),\ }\href {\doibase
  10.1103/PhysRevLett.109.101802} {\bibfield  {journal} {\bibinfo  {journal}
  {Phys. Rev. Lett.}\ }\textbf {\bibinfo {volume} {109}},\ \bibinfo {pages}
  {101802} (\bibinfo {year} {2012})},\ \Eprint {http://arxiv.org/abs/1205.5442}
  {arXiv:1205.5442 [hep-ex]} \BibitemShut {NoStop}%
\bibitem [{\citenamefont {Lees}\ \emph {et~al.}(2013)\citenamefont {Lees} \emph
  {et~al.}}]{Lees:2013uzd}%
  \BibitemOpen
  \bibfield  {author} {\bibinfo {author} {\bibfnamefont {J.}~\bibnamefont
  {Lees}} \emph {et~al.} (\bibinfo {collaboration} {BaBar}),\ }\href {\doibase
  10.1103/PhysRevD.88.072012} {\bibfield  {journal} {\bibinfo  {journal} {Phys.
  Rev. D}\ }\textbf {\bibinfo {volume} {88}},\ \bibinfo {pages} {072012}
  (\bibinfo {year} {2013})},\ \Eprint {http://arxiv.org/abs/1303.0571}
  {arXiv:1303.0571 [hep-ex]} \BibitemShut {NoStop}%
\bibitem [{\citenamefont {Aaij}\ \emph
  {et~al.}(2015{\natexlab{b}})\citenamefont {Aaij} \emph
  {et~al.}}]{Aaij:2015yra}%
  \BibitemOpen
  \bibfield  {author} {\bibinfo {author} {\bibfnamefont {R.}~\bibnamefont
  {Aaij}} \emph {et~al.} (\bibinfo {collaboration} {LHCb}),\ }\href {\doibase
  10.1103/PhysRevLett.115.111803} {\bibfield  {journal} {\bibinfo  {journal}
  {Phys. Rev. Lett.}\ }\textbf {\bibinfo {volume} {115}},\ \bibinfo {pages}
  {111803} (\bibinfo {year} {2015}{\natexlab{b}})},\ \bibinfo {note} {[Erratum:
  Phys.Rev.Lett. 115, 159901 (2015)]},\ \Eprint
  {http://arxiv.org/abs/1506.08614} {arXiv:1506.08614 [hep-ex]} \BibitemShut
  {NoStop}%
\bibitem [{\citenamefont {Aaij}\ \emph
  {et~al.}(2018{\natexlab{a}})\citenamefont {Aaij} \emph
  {et~al.}}]{Aaij:2017deq}%
  \BibitemOpen
  \bibfield  {author} {\bibinfo {author} {\bibfnamefont {R.}~\bibnamefont
  {Aaij}} \emph {et~al.} (\bibinfo {collaboration} {LHCb}),\ }\href {\doibase
  10.1103/PhysRevD.97.072013} {\bibfield  {journal} {\bibinfo  {journal} {Phys.
  Rev. D}\ }\textbf {\bibinfo {volume} {97}},\ \bibinfo {pages} {072013}
  (\bibinfo {year} {2018}{\natexlab{a}})},\ \Eprint
  {http://arxiv.org/abs/1711.02505} {arXiv:1711.02505 [hep-ex]} \BibitemShut
  {NoStop}%
\bibitem [{\citenamefont {Aaij}\ \emph
  {et~al.}(2018{\natexlab{b}})\citenamefont {Aaij} \emph
  {et~al.}}]{Aaij:2017uff}%
  \BibitemOpen
  \bibfield  {author} {\bibinfo {author} {\bibfnamefont {R.}~\bibnamefont
  {Aaij}} \emph {et~al.} (\bibinfo {collaboration} {LHCb}),\ }\href {\doibase
  10.1103/PhysRevLett.120.171802} {\bibfield  {journal} {\bibinfo  {journal}
  {Phys. Rev. Lett.}\ }\textbf {\bibinfo {volume} {120}},\ \bibinfo {pages}
  {171802} (\bibinfo {year} {2018}{\natexlab{b}})},\ \Eprint
  {http://arxiv.org/abs/1708.08856} {arXiv:1708.08856 [hep-ex]} \BibitemShut
  {NoStop}%
\bibitem [{\citenamefont {Abdesselam}\ \emph {et~al.}(2019)\citenamefont
  {Abdesselam} \emph {et~al.}}]{Abdesselam:2019dgh}%
  \BibitemOpen
  \bibfield  {author} {\bibinfo {author} {\bibfnamefont {A.}~\bibnamefont
  {Abdesselam}} \emph {et~al.} (\bibinfo {collaboration} {Belle}),\ }\href@noop
  {} {\  (\bibinfo {year} {2019})},\ \Eprint {http://arxiv.org/abs/1904.08794}
  {arXiv:1904.08794 [hep-ex]} \BibitemShut {NoStop}%
\bibitem [{\citenamefont {Capdevila}\ \emph {et~al.}(2018)\citenamefont
  {Capdevila}, \citenamefont {Crivellin}, \citenamefont {Descotes-Genon},
  \citenamefont {Matias},\ and\ \citenamefont {Virto}}]{Capdevila:2017bsm}%
  \BibitemOpen
  \bibfield  {author} {\bibinfo {author} {\bibfnamefont {B.}~\bibnamefont
  {Capdevila}}, \bibinfo {author} {\bibfnamefont {A.}~\bibnamefont
  {Crivellin}}, \bibinfo {author} {\bibfnamefont {S.}~\bibnamefont
  {Descotes-Genon}}, \bibinfo {author} {\bibfnamefont {J.}~\bibnamefont
  {Matias}}, \ and\ \bibinfo {author} {\bibfnamefont {J.}~\bibnamefont
  {Virto}},\ }\href {\doibase 10.1007/JHEP01(2018)093} {\bibfield  {journal}
  {\bibinfo  {journal} {JHEP}\ }\textbf {\bibinfo {volume} {01}},\ \bibinfo
  {pages} {093} (\bibinfo {year} {2018})},\ \Eprint
  {http://arxiv.org/abs/1704.05340} {arXiv:1704.05340 [hep-ph]} \BibitemShut
  {NoStop}%
\bibitem [{\citenamefont {Altmannshofer}\ \emph {et~al.}(2017)\citenamefont
  {Altmannshofer}, \citenamefont {Stangl},\ and\ \citenamefont
  {Straub}}]{Altmannshofer:2017yso}%
  \BibitemOpen
  \bibfield  {author} {\bibinfo {author} {\bibfnamefont {W.}~\bibnamefont
  {Altmannshofer}}, \bibinfo {author} {\bibfnamefont {P.}~\bibnamefont
  {Stangl}}, \ and\ \bibinfo {author} {\bibfnamefont {D.~M.}\ \bibnamefont
  {Straub}},\ }\href {\doibase 10.1103/PhysRevD.96.055008} {\bibfield
  {journal} {\bibinfo  {journal} {Phys. Rev. D}\ }\textbf {\bibinfo {volume}
  {96}},\ \bibinfo {pages} {055008} (\bibinfo {year} {2017})},\ \Eprint
  {http://arxiv.org/abs/1704.05435} {arXiv:1704.05435 [hep-ph]} \BibitemShut
  {NoStop}%
\bibitem [{\citenamefont {D'Amico}\ \emph {et~al.}(2017)\citenamefont
  {D'Amico}, \citenamefont {Nardecchia}, \citenamefont {Panci}, \citenamefont
  {Sannino}, \citenamefont {Strumia}, \citenamefont {Torre},\ and\
  \citenamefont {Urbano}}]{DAmico:2017mtc}%
  \BibitemOpen
  \bibfield  {author} {\bibinfo {author} {\bibfnamefont {G.}~\bibnamefont
  {D'Amico}}, \bibinfo {author} {\bibfnamefont {M.}~\bibnamefont {Nardecchia}},
  \bibinfo {author} {\bibfnamefont {P.}~\bibnamefont {Panci}}, \bibinfo
  {author} {\bibfnamefont {F.}~\bibnamefont {Sannino}}, \bibinfo {author}
  {\bibfnamefont {A.}~\bibnamefont {Strumia}}, \bibinfo {author} {\bibfnamefont
  {R.}~\bibnamefont {Torre}}, \ and\ \bibinfo {author} {\bibfnamefont
  {A.}~\bibnamefont {Urbano}},\ }\href {\doibase 10.1007/JHEP09(2017)010}
  {\bibfield  {journal} {\bibinfo  {journal} {JHEP}\ }\textbf {\bibinfo
  {volume} {09}},\ \bibinfo {pages} {010} (\bibinfo {year} {2017})},\ \Eprint
  {http://arxiv.org/abs/1704.05438} {arXiv:1704.05438 [hep-ph]} \BibitemShut
  {NoStop}%
\bibitem [{\citenamefont {Ciuchini}\ \emph {et~al.}(2017)\citenamefont
  {Ciuchini}, \citenamefont {Coutinho}, \citenamefont {Fedele}, \citenamefont
  {Franco}, \citenamefont {Paul}, \citenamefont {Silvestrini},\ and\
  \citenamefont {Valli}}]{Ciuchini:2017mik}%
  \BibitemOpen
  \bibfield  {author} {\bibinfo {author} {\bibfnamefont {M.}~\bibnamefont
  {Ciuchini}}, \bibinfo {author} {\bibfnamefont {A.~M.}\ \bibnamefont
  {Coutinho}}, \bibinfo {author} {\bibfnamefont {M.}~\bibnamefont {Fedele}},
  \bibinfo {author} {\bibfnamefont {E.}~\bibnamefont {Franco}}, \bibinfo
  {author} {\bibfnamefont {A.}~\bibnamefont {Paul}}, \bibinfo {author}
  {\bibfnamefont {L.}~\bibnamefont {Silvestrini}}, \ and\ \bibinfo {author}
  {\bibfnamefont {M.}~\bibnamefont {Valli}},\ }\href {\doibase
  10.1140/epjc/s10052-017-5270-2} {\bibfield  {journal} {\bibinfo  {journal}
  {Eur. Phys. J. C}\ }\textbf {\bibinfo {volume} {77}},\ \bibinfo {pages} {688}
  (\bibinfo {year} {2017})},\ \Eprint {http://arxiv.org/abs/1704.05447}
  {arXiv:1704.05447 [hep-ph]} \BibitemShut {NoStop}%
\bibitem [{\citenamefont {Hiller}\ and\ \citenamefont
  {Nisandzic}(2017)}]{Hiller:2017bzc}%
  \BibitemOpen
  \bibfield  {author} {\bibinfo {author} {\bibfnamefont {G.}~\bibnamefont
  {Hiller}}\ and\ \bibinfo {author} {\bibfnamefont {I.}~\bibnamefont
  {Nisandzic}},\ }\href {\doibase 10.1103/PhysRevD.96.035003} {\bibfield
  {journal} {\bibinfo  {journal} {Phys. Rev. D}\ }\textbf {\bibinfo {volume}
  {96}},\ \bibinfo {pages} {035003} (\bibinfo {year} {2017})},\ \Eprint
  {http://arxiv.org/abs/1704.05444} {arXiv:1704.05444 [hep-ph]} \BibitemShut
  {NoStop}%
\bibitem [{\citenamefont {Geng}\ \emph {et~al.}(2017)\citenamefont {Geng},
  \citenamefont {Grinstein}, \citenamefont {J\"ager}, \citenamefont
  {Martin~Camalich}, \citenamefont {Ren},\ and\ \citenamefont
  {Shi}}]{Geng:2017svp}%
  \BibitemOpen
  \bibfield  {author} {\bibinfo {author} {\bibfnamefont {L.-S.}\ \bibnamefont
  {Geng}}, \bibinfo {author} {\bibfnamefont {B.}~\bibnamefont {Grinstein}},
  \bibinfo {author} {\bibfnamefont {S.}~\bibnamefont {J\"ager}}, \bibinfo
  {author} {\bibfnamefont {J.}~\bibnamefont {Martin~Camalich}}, \bibinfo
  {author} {\bibfnamefont {X.-L.}\ \bibnamefont {Ren}}, \ and\ \bibinfo
  {author} {\bibfnamefont {R.-X.}\ \bibnamefont {Shi}},\ }\href {\doibase
  10.1103/PhysRevD.96.093006} {\bibfield  {journal} {\bibinfo  {journal} {Phys.
  Rev. D}\ }\textbf {\bibinfo {volume} {96}},\ \bibinfo {pages} {093006}
  (\bibinfo {year} {2017})},\ \Eprint {http://arxiv.org/abs/1704.05446}
  {arXiv:1704.05446 [hep-ph]} \BibitemShut {NoStop}%
\bibitem [{\citenamefont {Hurth}\ \emph {et~al.}(2017)\citenamefont {Hurth},
  \citenamefont {Mahmoudi}, \citenamefont {Martinez~Santos},\ and\
  \citenamefont {Neshatpour}}]{Hurth:2017hxg}%
  \BibitemOpen
  \bibfield  {author} {\bibinfo {author} {\bibfnamefont {T.}~\bibnamefont
  {Hurth}}, \bibinfo {author} {\bibfnamefont {F.}~\bibnamefont {Mahmoudi}},
  \bibinfo {author} {\bibfnamefont {D.}~\bibnamefont {Martinez~Santos}}, \ and\
  \bibinfo {author} {\bibfnamefont {S.}~\bibnamefont {Neshatpour}},\ }\href
  {\doibase 10.1103/PhysRevD.96.095034} {\bibfield  {journal} {\bibinfo
  {journal} {Phys. Rev. D}\ }\textbf {\bibinfo {volume} {96}},\ \bibinfo
  {pages} {095034} (\bibinfo {year} {2017})},\ \Eprint
  {http://arxiv.org/abs/1705.06274} {arXiv:1705.06274 [hep-ph]} \BibitemShut
  {NoStop}%
\bibitem [{\citenamefont {Alok}\ \emph {et~al.}(2017)\citenamefont {Alok},
  \citenamefont {Bhattacharya}, \citenamefont {Datta}, \citenamefont {Kumar},
  \citenamefont {Kumar},\ and\ \citenamefont {London}}]{Alok:2017sui}%
  \BibitemOpen
  \bibfield  {author} {\bibinfo {author} {\bibfnamefont {A.~K.}\ \bibnamefont
  {Alok}}, \bibinfo {author} {\bibfnamefont {B.}~\bibnamefont {Bhattacharya}},
  \bibinfo {author} {\bibfnamefont {A.}~\bibnamefont {Datta}}, \bibinfo
  {author} {\bibfnamefont {D.}~\bibnamefont {Kumar}}, \bibinfo {author}
  {\bibfnamefont {J.}~\bibnamefont {Kumar}}, \ and\ \bibinfo {author}
  {\bibfnamefont {D.}~\bibnamefont {London}},\ }\href {\doibase
  10.1103/PhysRevD.96.095009} {\bibfield  {journal} {\bibinfo  {journal} {Phys.
  Rev. D}\ }\textbf {\bibinfo {volume} {96}},\ \bibinfo {pages} {095009}
  (\bibinfo {year} {2017})},\ \Eprint {http://arxiv.org/abs/1704.07397}
  {arXiv:1704.07397 [hep-ph]} \BibitemShut {NoStop}%
\bibitem [{\citenamefont {Alguer\'o}\ \emph {et~al.}(2019)\citenamefont
  {Alguer\'o}, \citenamefont {Capdevila}, \citenamefont {Crivellin},
  \citenamefont {Descotes-Genon}, \citenamefont {Masjuan}, \citenamefont
  {Matias}, \citenamefont {Novoa~Brunet},\ and\ \citenamefont
  {Virto}}]{Alguero:2019ptt}%
  \BibitemOpen
  \bibfield  {author} {\bibinfo {author} {\bibfnamefont {M.}~\bibnamefont
  {Alguer\'o}}, \bibinfo {author} {\bibfnamefont {B.}~\bibnamefont
  {Capdevila}}, \bibinfo {author} {\bibfnamefont {A.}~\bibnamefont
  {Crivellin}}, \bibinfo {author} {\bibfnamefont {S.}~\bibnamefont
  {Descotes-Genon}}, \bibinfo {author} {\bibfnamefont {P.}~\bibnamefont
  {Masjuan}}, \bibinfo {author} {\bibfnamefont {J.}~\bibnamefont {Matias}},
  \bibinfo {author} {\bibfnamefont {M.}~\bibnamefont {Novoa~Brunet}}, \ and\
  \bibinfo {author} {\bibfnamefont {J.}~\bibnamefont {Virto}},\ }\href
  {\doibase 10.1140/epjc/s10052-019-7216-3} {\bibfield  {journal} {\bibinfo
  {journal} {Eur. Phys. J. C}\ }\textbf {\bibinfo {volume} {79}},\ \bibinfo
  {pages} {714} (\bibinfo {year} {2019})},\ \bibinfo {note} {[Addendum:
  Eur.Phys.J.C 80, 511 (2020)]},\ \Eprint {http://arxiv.org/abs/1903.09578}
  {arXiv:1903.09578 [hep-ph]} \BibitemShut {NoStop}%
\bibitem [{\citenamefont {Aebischer}\ \emph {et~al.}(2020)\citenamefont
  {Aebischer}, \citenamefont {Altmannshofer}, \citenamefont {Guadagnoli},
  \citenamefont {Reboud}, \citenamefont {Stangl},\ and\ \citenamefont
  {Straub}}]{Aebischer:2019mlg}%
  \BibitemOpen
  \bibfield  {author} {\bibinfo {author} {\bibfnamefont {J.}~\bibnamefont
  {Aebischer}}, \bibinfo {author} {\bibfnamefont {W.}~\bibnamefont
  {Altmannshofer}}, \bibinfo {author} {\bibfnamefont {D.}~\bibnamefont
  {Guadagnoli}}, \bibinfo {author} {\bibfnamefont {M.}~\bibnamefont {Reboud}},
  \bibinfo {author} {\bibfnamefont {P.}~\bibnamefont {Stangl}}, \ and\ \bibinfo
  {author} {\bibfnamefont {D.~M.}\ \bibnamefont {Straub}},\ }\href {\doibase
  10.1140/epjc/s10052-020-7817-x} {\bibfield  {journal} {\bibinfo  {journal}
  {Eur. Phys. J. C}\ }\textbf {\bibinfo {volume} {80}},\ \bibinfo {pages} {252}
  (\bibinfo {year} {2020})},\ \Eprint {http://arxiv.org/abs/1903.10434}
  {arXiv:1903.10434 [hep-ph]} \BibitemShut {NoStop}%
\bibitem [{\citenamefont {Ciuchini}\ \emph {et~al.}(2019)\citenamefont
  {Ciuchini}, \citenamefont {Coutinho}, \citenamefont {Fedele}, \citenamefont
  {Franco}, \citenamefont {Paul}, \citenamefont {Silvestrini},\ and\
  \citenamefont {Valli}}]{Ciuchini:2019usw}%
  \BibitemOpen
  \bibfield  {author} {\bibinfo {author} {\bibfnamefont {M.}~\bibnamefont
  {Ciuchini}}, \bibinfo {author} {\bibfnamefont {A.~M.}\ \bibnamefont
  {Coutinho}}, \bibinfo {author} {\bibfnamefont {M.}~\bibnamefont {Fedele}},
  \bibinfo {author} {\bibfnamefont {E.}~\bibnamefont {Franco}}, \bibinfo
  {author} {\bibfnamefont {A.}~\bibnamefont {Paul}}, \bibinfo {author}
  {\bibfnamefont {L.}~\bibnamefont {Silvestrini}}, \ and\ \bibinfo {author}
  {\bibfnamefont {M.}~\bibnamefont {Valli}},\ }\href {\doibase
  10.1140/epjc/s10052-019-7210-9} {\bibfield  {journal} {\bibinfo  {journal}
  {Eur. Phys. J. C}\ }\textbf {\bibinfo {volume} {79}},\ \bibinfo {pages} {719}
  (\bibinfo {year} {2019})},\ \Eprint {http://arxiv.org/abs/1903.09632}
  {arXiv:1903.09632 [hep-ph]} \BibitemShut {NoStop}%
\bibitem [{\citenamefont {Ciuchini}\ \emph {et~al.}(2020)\citenamefont
  {Ciuchini}, \citenamefont {Fedele}, \citenamefont {Franco}, \citenamefont
  {Paul}, \citenamefont {Silvestrini},\ and\ \citenamefont
  {Valli}}]{Ciuchini:2020gvn}%
  \BibitemOpen
  \bibfield  {author} {\bibinfo {author} {\bibfnamefont {M.}~\bibnamefont
  {Ciuchini}}, \bibinfo {author} {\bibfnamefont {M.}~\bibnamefont {Fedele}},
  \bibinfo {author} {\bibfnamefont {E.}~\bibnamefont {Franco}}, \bibinfo
  {author} {\bibfnamefont {A.}~\bibnamefont {Paul}}, \bibinfo {author}
  {\bibfnamefont {L.}~\bibnamefont {Silvestrini}}, \ and\ \bibinfo {author}
  {\bibfnamefont {M.}~\bibnamefont {Valli}},\ }\href@noop {} {\  (\bibinfo
  {year} {2020})},\ \Eprint {http://arxiv.org/abs/2011.01212} {arXiv:2011.01212
  [hep-ph]} \BibitemShut {NoStop}%
\bibitem [{\citenamefont {Amhis}\ \emph {et~al.}(2017)\citenamefont {Amhis}
  \emph {et~al.}}]{Amhis:2016xyh}%
  \BibitemOpen
  \bibfield  {author} {\bibinfo {author} {\bibfnamefont {Y.}~\bibnamefont
  {Amhis}} \emph {et~al.} (\bibinfo {collaboration} {HFLAV}),\ }\href {\doibase
  10.1140/epjc/s10052-017-5058-4} {\bibfield  {journal} {\bibinfo  {journal}
  {Eur. Phys. J. C}\ }\textbf {\bibinfo {volume} {77}},\ \bibinfo {pages} {895}
  (\bibinfo {year} {2017})},\ \Eprint {http://arxiv.org/abs/1612.07233}
  {arXiv:1612.07233 [hep-ex]} \BibitemShut {NoStop}%
\bibitem [{\citenamefont {Murgui}\ \emph {et~al.}(2019)\citenamefont {Murgui},
  \citenamefont {Pe\~nuelas}, \citenamefont {Jung},\ and\ \citenamefont
  {Pich}}]{Murgui:2019czp}%
  \BibitemOpen
  \bibfield  {author} {\bibinfo {author} {\bibfnamefont {C.}~\bibnamefont
  {Murgui}}, \bibinfo {author} {\bibfnamefont {A.}~\bibnamefont {Pe\~nuelas}},
  \bibinfo {author} {\bibfnamefont {M.}~\bibnamefont {Jung}}, \ and\ \bibinfo
  {author} {\bibfnamefont {A.}~\bibnamefont {Pich}},\ }\href {\doibase
  10.1007/JHEP09(2019)103} {\bibfield  {journal} {\bibinfo  {journal} {JHEP}\
  }\textbf {\bibinfo {volume} {09}},\ \bibinfo {pages} {103} (\bibinfo {year}
  {2019})},\ \Eprint {http://arxiv.org/abs/1904.09311} {arXiv:1904.09311
  [hep-ph]} \BibitemShut {NoStop}%
\bibitem [{\citenamefont {Shi}\ \emph {et~al.}(2019)\citenamefont {Shi},
  \citenamefont {Geng}, \citenamefont {Grinstein}, \citenamefont {J\"ager},\
  and\ \citenamefont {Martin~Camalich}}]{Shi:2019gxi}%
  \BibitemOpen
  \bibfield  {author} {\bibinfo {author} {\bibfnamefont {R.-X.}\ \bibnamefont
  {Shi}}, \bibinfo {author} {\bibfnamefont {L.-S.}\ \bibnamefont {Geng}},
  \bibinfo {author} {\bibfnamefont {B.}~\bibnamefont {Grinstein}}, \bibinfo
  {author} {\bibfnamefont {S.}~\bibnamefont {J\"ager}}, \ and\ \bibinfo
  {author} {\bibfnamefont {J.}~\bibnamefont {Martin~Camalich}},\ }\href
  {\doibase 10.1007/JHEP12(2019)065} {\bibfield  {journal} {\bibinfo  {journal}
  {JHEP}\ }\textbf {\bibinfo {volume} {12}},\ \bibinfo {pages} {065} (\bibinfo
  {year} {2019})},\ \Eprint {http://arxiv.org/abs/1905.08498} {arXiv:1905.08498
  [hep-ph]} \BibitemShut {NoStop}%
\bibitem [{\citenamefont {Blanke}\ \emph {et~al.}(2019)\citenamefont {Blanke},
  \citenamefont {Crivellin}, \citenamefont {Kitahara}, \citenamefont {Moscati},
  \citenamefont {Nierste},\ and\ \citenamefont
  {Ni\v{s}and\v{z}i\'c}}]{Blanke:2019qrx}%
  \BibitemOpen
  \bibfield  {author} {\bibinfo {author} {\bibfnamefont {M.}~\bibnamefont
  {Blanke}}, \bibinfo {author} {\bibfnamefont {A.}~\bibnamefont {Crivellin}},
  \bibinfo {author} {\bibfnamefont {T.}~\bibnamefont {Kitahara}}, \bibinfo
  {author} {\bibfnamefont {M.}~\bibnamefont {Moscati}}, \bibinfo {author}
  {\bibfnamefont {U.}~\bibnamefont {Nierste}}, \ and\ \bibinfo {author}
  {\bibfnamefont {I.}~\bibnamefont {Ni\v{s}and\v{z}i\'c}},\ }\href {\doibase
  10.1103/PhysRevD.100.035035} {\  (\bibinfo {year} {2019}),\
  10.1103/PhysRevD.100.035035},\ \bibinfo {note} {[Addendum: Phys.Rev.D 100,
  035035 (2019)]},\ \Eprint {http://arxiv.org/abs/1905.08253} {arXiv:1905.08253
  [hep-ph]} \BibitemShut {NoStop}%
\bibitem [{\citenamefont {Kumbhakar}\ \emph {et~al.}(2020)\citenamefont
  {Kumbhakar}, \citenamefont {Alok}, \citenamefont {Kumar},\ and\ \citenamefont
  {Sankar}}]{Kumbhakar:2019avh}%
  \BibitemOpen
  \bibfield  {author} {\bibinfo {author} {\bibfnamefont {S.}~\bibnamefont
  {Kumbhakar}}, \bibinfo {author} {\bibfnamefont {A.~K.}\ \bibnamefont {Alok}},
  \bibinfo {author} {\bibfnamefont {D.}~\bibnamefont {Kumar}}, \ and\ \bibinfo
  {author} {\bibfnamefont {S.~U.}\ \bibnamefont {Sankar}},\ }\href {\doibase
  10.22323/1.364.0272} {\bibfield  {journal} {\bibinfo  {journal} {PoS}\
  }\textbf {\bibinfo {volume} {EPS-HEP2019}},\ \bibinfo {pages} {272} (\bibinfo
  {year} {2020})},\ \Eprint {http://arxiv.org/abs/1909.02840} {arXiv:1909.02840
  [hep-ph]} \BibitemShut {NoStop}%
\bibitem [{\citenamefont {Bennett}\ \emph {et~al.}(2006)\citenamefont {Bennett}
  \emph {et~al.}}]{Bennett:2006fi}%
  \BibitemOpen
  \bibfield  {author} {\bibinfo {author} {\bibfnamefont {G.}~\bibnamefont
  {Bennett}} \emph {et~al.} (\bibinfo {collaboration} {Muon g-2}),\ }\href
  {\doibase 10.1103/PhysRevD.73.072003} {\bibfield  {journal} {\bibinfo
  {journal} {Phys. Rev. D}\ }\textbf {\bibinfo {volume} {73}},\ \bibinfo
  {pages} {072003} (\bibinfo {year} {2006})},\ \Eprint
  {http://arxiv.org/abs/hep-ex/0602035} {arXiv:hep-ex/0602035} \BibitemShut
  {NoStop}%
\bibitem [{\citenamefont {Aoyama}\ \emph {et~al.}(2020)\citenamefont {Aoyama}
  \emph {et~al.}}]{Aoyama:2020ynm}%
  \BibitemOpen
  \bibfield  {author} {\bibinfo {author} {\bibfnamefont {T.}~\bibnamefont
  {Aoyama}} \emph {et~al.},\ }\href {\doibase 10.1016/j.physrep.2020.07.006}
  {\bibfield  {journal} {\bibinfo  {journal} {Phys. Rept.}\ }\textbf {\bibinfo
  {volume} {887}},\ \bibinfo {pages} {1} (\bibinfo {year} {2020})},\ \Eprint
  {http://arxiv.org/abs/2006.04822} {arXiv:2006.04822 [hep-ph]} \BibitemShut
  {NoStop}%
\bibitem [{\citenamefont {Belfatto}\ \emph {et~al.}(2020)\citenamefont
  {Belfatto}, \citenamefont {Beradze},\ and\ \citenamefont
  {Berezhiani}}]{Belfatto:2019swo}%
  \BibitemOpen
  \bibfield  {author} {\bibinfo {author} {\bibfnamefont {B.}~\bibnamefont
  {Belfatto}}, \bibinfo {author} {\bibfnamefont {R.}~\bibnamefont {Beradze}}, \
  and\ \bibinfo {author} {\bibfnamefont {Z.}~\bibnamefont {Berezhiani}},\
  }\href {\doibase 10.1140/epjc/s10052-020-7691-6} {\bibfield  {journal}
  {\bibinfo  {journal} {Eur. Phys. J. C}\ }\textbf {\bibinfo {volume} {80}},\
  \bibinfo {pages} {149} (\bibinfo {year} {2020})},\ \Eprint
  {http://arxiv.org/abs/1906.02714} {arXiv:1906.02714 [hep-ph]} \BibitemShut
  {NoStop}%
\bibitem [{\citenamefont {Grossman}\ \emph {et~al.}(2020)\citenamefont
  {Grossman}, \citenamefont {Passemar},\ and\ \citenamefont
  {Schacht}}]{Grossman:2019bzp}%
  \BibitemOpen
  \bibfield  {author} {\bibinfo {author} {\bibfnamefont {Y.}~\bibnamefont
  {Grossman}}, \bibinfo {author} {\bibfnamefont {E.}~\bibnamefont {Passemar}},
  \ and\ \bibinfo {author} {\bibfnamefont {S.}~\bibnamefont {Schacht}},\ }\href
  {\doibase 10.1007/JHEP07(2020)068} {\bibfield  {journal} {\bibinfo  {journal}
  {JHEP}\ }\textbf {\bibinfo {volume} {07}},\ \bibinfo {pages} {068} (\bibinfo
  {year} {2020})},\ \Eprint {http://arxiv.org/abs/1911.07821} {arXiv:1911.07821
  [hep-ph]} \BibitemShut {NoStop}%
\bibitem [{\citenamefont {Coutinho}\ \emph {et~al.}(2020)\citenamefont
  {Coutinho}, \citenamefont {Crivellin},\ and\ \citenamefont
  {Manzari}}]{Coutinho:2019aiy}%
  \BibitemOpen
  \bibfield  {author} {\bibinfo {author} {\bibfnamefont {A.~M.}\ \bibnamefont
  {Coutinho}}, \bibinfo {author} {\bibfnamefont {A.}~\bibnamefont {Crivellin}},
  \ and\ \bibinfo {author} {\bibfnamefont {C.~A.}\ \bibnamefont {Manzari}},\
  }\href {\doibase 10.1103/PhysRevLett.125.071802} {\bibfield  {journal}
  {\bibinfo  {journal} {Phys. Rev. Lett.}\ }\textbf {\bibinfo {volume} {125}},\
  \bibinfo {pages} {071802} (\bibinfo {year} {2020})},\ \Eprint
  {http://arxiv.org/abs/1912.08823} {arXiv:1912.08823 [hep-ph]} \BibitemShut
  {NoStop}%
\bibitem [{\citenamefont {Crivellin}\ and\ \citenamefont
  {Hoferichter}(2020)}]{Crivellin:2020lzu}%
  \BibitemOpen
  \bibfield  {author} {\bibinfo {author} {\bibfnamefont {A.}~\bibnamefont
  {Crivellin}}\ and\ \bibinfo {author} {\bibfnamefont {M.}~\bibnamefont
  {Hoferichter}},\ }\href {\doibase 10.1103/PhysRevLett.125.111801} {\bibfield
  {journal} {\bibinfo  {journal} {Phys. Rev. Lett.}\ }\textbf {\bibinfo
  {volume} {125}},\ \bibinfo {pages} {111801} (\bibinfo {year} {2020})},\
  \Eprint {http://arxiv.org/abs/2002.07184} {arXiv:2002.07184 [hep-ph]}
  \BibitemShut {NoStop}%
\bibitem [{\citenamefont {Manzari}\ \emph {et~al.}(2021)\citenamefont
  {Manzari}, \citenamefont {Coutinho},\ and\ \citenamefont
  {Crivellin}}]{Coutinho:2020xhc}%
  \BibitemOpen
  \bibfield  {author} {\bibinfo {author} {\bibfnamefont {C.~A.}\ \bibnamefont
  {Manzari}}, \bibinfo {author} {\bibfnamefont {A.~M.}\ \bibnamefont
  {Coutinho}}, \ and\ \bibinfo {author} {\bibfnamefont {A.}~\bibnamefont
  {Crivellin}},\ }\href {\doibase 10.22323/1.382.0242} {\bibfield  {journal}
  {\bibinfo  {journal} {PoS}\ }\textbf {\bibinfo {volume} {LHCP2020}},\
  \bibinfo {pages} {242} (\bibinfo {year} {2021})},\ \Eprint
  {http://arxiv.org/abs/2009.03877} {arXiv:2009.03877 [hep-ph]} \BibitemShut
  {NoStop}%
\bibitem [{\citenamefont {Crivellin}\ \emph
  {et~al.}(2020{\natexlab{a}})\citenamefont {Crivellin}, \citenamefont {Kirk},
  \citenamefont {Manzari},\ and\ \citenamefont {Montull}}]{Crivellin:2020ebi}%
  \BibitemOpen
  \bibfield  {author} {\bibinfo {author} {\bibfnamefont {A.}~\bibnamefont
  {Crivellin}}, \bibinfo {author} {\bibfnamefont {F.}~\bibnamefont {Kirk}},
  \bibinfo {author} {\bibfnamefont {C.~A.}\ \bibnamefont {Manzari}}, \ and\
  \bibinfo {author} {\bibfnamefont {M.}~\bibnamefont {Montull}},\ }\href@noop
  {} {\  (\bibinfo {year} {2020}{\natexlab{a}})},\ \Eprint
  {http://arxiv.org/abs/2008.01113} {arXiv:2008.01113 [hep-ph]} \BibitemShut
  {NoStop}%
\bibitem [{\citenamefont {Kirk}(2020)}]{Kirk:2020wdk}%
  \BibitemOpen
  \bibfield  {author} {\bibinfo {author} {\bibfnamefont {M.}~\bibnamefont
  {Kirk}},\ }\href@noop {} {\  (\bibinfo {year} {2020})},\ \Eprint
  {http://arxiv.org/abs/2008.03261} {arXiv:2008.03261 [hep-ph]} \BibitemShut
  {NoStop}%
\bibitem [{\citenamefont {Alok}\ \emph {et~al.}(2020)\citenamefont {Alok},
  \citenamefont {Dighe}, \citenamefont {Gangal},\ and\ \citenamefont
  {Kumar}}]{Alok:2020jod}%
  \BibitemOpen
  \bibfield  {author} {\bibinfo {author} {\bibfnamefont {A.~K.}\ \bibnamefont
  {Alok}}, \bibinfo {author} {\bibfnamefont {A.}~\bibnamefont {Dighe}},
  \bibinfo {author} {\bibfnamefont {S.}~\bibnamefont {Gangal}}, \ and\ \bibinfo
  {author} {\bibfnamefont {J.}~\bibnamefont {Kumar}},\ }\href@noop {} {\
  (\bibinfo {year} {2020})},\ \Eprint {http://arxiv.org/abs/2010.12009}
  {arXiv:2010.12009 [hep-ph]} \BibitemShut {NoStop}%
\bibitem [{\citenamefont {Crivellin}\ \emph
  {et~al.}(2020{\natexlab{b}})\citenamefont {Crivellin}, \citenamefont
  {Manzari}, \citenamefont {Alguero},\ and\ \citenamefont
  {Matias}}]{Crivellin:2020oup}%
  \BibitemOpen
  \bibfield  {author} {\bibinfo {author} {\bibfnamefont {A.}~\bibnamefont
  {Crivellin}}, \bibinfo {author} {\bibfnamefont {C.~A.}\ \bibnamefont
  {Manzari}}, \bibinfo {author} {\bibfnamefont {M.}~\bibnamefont {Alguero}}, \
  and\ \bibinfo {author} {\bibfnamefont {J.}~\bibnamefont {Matias}},\
  }\href@noop {} {\  (\bibinfo {year} {2020}{\natexlab{b}})},\ \Eprint
  {http://arxiv.org/abs/2010.14504} {arXiv:2010.14504 [hep-ph]} \BibitemShut
  {NoStop}%
\bibitem [{\citenamefont {Shiells}\ \emph {et~al.}(2020)\citenamefont
  {Shiells}, \citenamefont {Blunden},\ and\ \citenamefont
  {Melnitchouk}}]{Shiells:2020fqp}%
  \BibitemOpen
  \bibfield  {author} {\bibinfo {author} {\bibfnamefont {K.}~\bibnamefont
  {Shiells}}, \bibinfo {author} {\bibfnamefont {P.}~\bibnamefont {Blunden}}, \
  and\ \bibinfo {author} {\bibfnamefont {W.}~\bibnamefont {Melnitchouk}},\
  }\href@noop {} {\  (\bibinfo {year} {2020})},\ \Eprint
  {http://arxiv.org/abs/2012.01580} {arXiv:2012.01580 [hep-ph]} \BibitemShut
  {NoStop}%
\bibitem [{\citenamefont {Seng}\ \emph {et~al.}(2020)\citenamefont {Seng},
  \citenamefont {Feng}, \citenamefont {Gorchtein},\ and\ \citenamefont
  {Jin}}]{Seng:2020wjq}%
  \BibitemOpen
  \bibfield  {author} {\bibinfo {author} {\bibfnamefont {C.-Y.}\ \bibnamefont
  {Seng}}, \bibinfo {author} {\bibfnamefont {X.}~\bibnamefont {Feng}}, \bibinfo
  {author} {\bibfnamefont {M.}~\bibnamefont {Gorchtein}}, \ and\ \bibinfo
  {author} {\bibfnamefont {L.-C.}\ \bibnamefont {Jin}},\ }\href {\doibase
  10.1103/PhysRevD.101.111301} {\bibfield  {journal} {\bibinfo  {journal}
  {Phys. Rev. D}\ }\textbf {\bibinfo {volume} {101}},\ \bibinfo {pages}
  {111301} (\bibinfo {year} {2020})},\ \Eprint
  {http://arxiv.org/abs/2003.11264} {arXiv:2003.11264 [hep-ph]} \BibitemShut
  {NoStop}%
\bibitem [{\citenamefont {Amhis}\ \emph {et~al.}(2019)\citenamefont {Amhis}
  \emph {et~al.}}]{Amhis:2019ckw}%
  \BibitemOpen
  \bibfield  {author} {\bibinfo {author} {\bibfnamefont {Y.~S.}\ \bibnamefont
  {Amhis}} \emph {et~al.} (\bibinfo {collaboration} {HFLAV}),\ }\href@noop {}
  {\  (\bibinfo {year} {2019})},\ \Eprint {http://arxiv.org/abs/1909.12524}
  {arXiv:1909.12524 [hep-ex]} \BibitemShut {NoStop}%
\bibitem [{\citenamefont {Capdevila}\ \emph {et~al.}(2020)\citenamefont
  {Capdevila}, \citenamefont {Crivellin}, \citenamefont {Manzari},\ and\
  \citenamefont {Montull}}]{Capdevila:2020rrl}%
  \BibitemOpen
  \bibfield  {author} {\bibinfo {author} {\bibfnamefont {B.}~\bibnamefont
  {Capdevila}}, \bibinfo {author} {\bibfnamefont {A.}~\bibnamefont
  {Crivellin}}, \bibinfo {author} {\bibfnamefont {C.~A.}\ \bibnamefont
  {Manzari}}, \ and\ \bibinfo {author} {\bibfnamefont {M.}~\bibnamefont
  {Montull}},\ }\href@noop {} {\  (\bibinfo {year} {2020})},\ \Eprint
  {http://arxiv.org/abs/2005.13542} {arXiv:2005.13542 [hep-ph]} \BibitemShut
  {NoStop}%
\bibitem [{\citenamefont {Buras}\ \emph {et~al.}()\citenamefont {Buras},
  \citenamefont {Crivellin}, \citenamefont {Kirk}, \citenamefont {Manzari},\
  and\ \citenamefont {Montull}}]{BCKMM}%
  \BibitemOpen
  \bibfield  {author} {\bibinfo {author} {\bibfnamefont {A.}~\bibnamefont
  {Buras}}, \bibinfo {author} {\bibfnamefont {A.}~\bibnamefont {Crivellin}},
  \bibinfo {author} {\bibfnamefont {F.}~\bibnamefont {Kirk}}, \bibinfo {author}
  {\bibfnamefont {C.~A.}\ \bibnamefont {Manzari}}, \ and\ \bibinfo {author}
  {\bibfnamefont {M.}~\bibnamefont {Montull}},\ }\href@noop {} {\ }\Eprint
  {http://arxiv.org/abs/In preparation} {In preparation} \BibitemShut {NoStop}%
\bibitem [{\citenamefont {Zee}(1986)}]{Zee:1985id}%
  \BibitemOpen
  \bibfield  {author} {\bibinfo {author} {\bibfnamefont {A.}~\bibnamefont
  {Zee}},\ }\href {\doibase 10.1016/0550-3213(86)90475-X} {\bibfield  {journal}
  {\bibinfo  {journal} {Nucl. Phys. B}\ }\textbf {\bibinfo {volume} {264}},\
  \bibinfo {pages} {99} (\bibinfo {year} {1986})}\BibitemShut {NoStop}%
\bibitem [{\citenamefont {Babu}(1988)}]{Babu:1988ki}%
  \BibitemOpen
  \bibfield  {author} {\bibinfo {author} {\bibfnamefont {K.}~\bibnamefont
  {Babu}},\ }\href {\doibase 10.1016/0370-2693(88)91584-5} {\bibfield
  {journal} {\bibinfo  {journal} {Phys. Lett. B}\ }\textbf {\bibinfo {volume}
  {203}},\ \bibinfo {pages} {132} (\bibinfo {year} {1988})}\BibitemShut
  {NoStop}%
\bibitem [{\citenamefont {Krauss}\ \emph {et~al.}(2003)\citenamefont {Krauss},
  \citenamefont {Nasri},\ and\ \citenamefont {Trodden}}]{Krauss:2002px}%
  \BibitemOpen
  \bibfield  {author} {\bibinfo {author} {\bibfnamefont {L.~M.}\ \bibnamefont
  {Krauss}}, \bibinfo {author} {\bibfnamefont {S.}~\bibnamefont {Nasri}}, \
  and\ \bibinfo {author} {\bibfnamefont {M.}~\bibnamefont {Trodden}},\ }\href
  {\doibase 10.1103/PhysRevD.67.085002} {\bibfield  {journal} {\bibinfo
  {journal} {Phys. Rev. D}\ }\textbf {\bibinfo {volume} {67}},\ \bibinfo
  {pages} {085002} (\bibinfo {year} {2003})},\ \Eprint
  {http://arxiv.org/abs/hep-ph/0210389} {arXiv:hep-ph/0210389} \BibitemShut
  {NoStop}%
\bibitem [{\citenamefont {Nebot}\ \emph {et~al.}(2008)\citenamefont {Nebot},
  \citenamefont {Oliver}, \citenamefont {Palao},\ and\ \citenamefont
  {Santamaria}}]{Nebot:2007bc}%
  \BibitemOpen
  \bibfield  {author} {\bibinfo {author} {\bibfnamefont {M.}~\bibnamefont
  {Nebot}}, \bibinfo {author} {\bibfnamefont {J.~F.}\ \bibnamefont {Oliver}},
  \bibinfo {author} {\bibfnamefont {D.}~\bibnamefont {Palao}}, \ and\ \bibinfo
  {author} {\bibfnamefont {A.}~\bibnamefont {Santamaria}},\ }\href {\doibase
  10.1103/PhysRevD.77.093013} {\bibfield  {journal} {\bibinfo  {journal} {Phys.
  Rev. D}\ }\textbf {\bibinfo {volume} {77}},\ \bibinfo {pages} {093013}
  (\bibinfo {year} {2008})},\ \Eprint {http://arxiv.org/abs/0711.0483}
  {arXiv:0711.0483 [hep-ph]} \BibitemShut {NoStop}%
\bibitem [{\citenamefont {Cai}\ \emph {et~al.}(2015)\citenamefont {Cai},
  \citenamefont {Clarke}, \citenamefont {Schmidt},\ and\ \citenamefont
  {Volkas}}]{Cai:2014kra}%
  \BibitemOpen
  \bibfield  {author} {\bibinfo {author} {\bibfnamefont {Y.}~\bibnamefont
  {Cai}}, \bibinfo {author} {\bibfnamefont {J.~D.}\ \bibnamefont {Clarke}},
  \bibinfo {author} {\bibfnamefont {M.~A.}\ \bibnamefont {Schmidt}}, \ and\
  \bibinfo {author} {\bibfnamefont {R.~R.}\ \bibnamefont {Volkas}},\ }\href
  {\doibase 10.1007/JHEP02(2015)161} {\bibfield  {journal} {\bibinfo  {journal}
  {JHEP}\ }\textbf {\bibinfo {volume} {02}},\ \bibinfo {pages} {161} (\bibinfo
  {year} {2015})},\ \Eprint {http://arxiv.org/abs/1410.0689} {arXiv:1410.0689
  [hep-ph]} \BibitemShut {NoStop}%
\bibitem [{\citenamefont {Cheung}\ and\ \citenamefont
  {Seto}(2004)}]{Cheung:2004xm}%
  \BibitemOpen
  \bibfield  {author} {\bibinfo {author} {\bibfnamefont {K.}~\bibnamefont
  {Cheung}}\ and\ \bibinfo {author} {\bibfnamefont {O.}~\bibnamefont {Seto}},\
  }\href {\doibase 10.1103/PhysRevD.69.113009} {\bibfield  {journal} {\bibinfo
  {journal} {Phys. Rev. D}\ }\textbf {\bibinfo {volume} {69}},\ \bibinfo
  {pages} {113009} (\bibinfo {year} {2004})},\ \Eprint
  {http://arxiv.org/abs/hep-ph/0403003} {arXiv:hep-ph/0403003} \BibitemShut
  {NoStop}%
\bibitem [{\citenamefont {Ahriche}\ \emph {et~al.}(2014)\citenamefont
  {Ahriche}, \citenamefont {Nasri},\ and\ \citenamefont
  {Soualah}}]{Ahriche:2014xra}%
  \BibitemOpen
  \bibfield  {author} {\bibinfo {author} {\bibfnamefont {A.}~\bibnamefont
  {Ahriche}}, \bibinfo {author} {\bibfnamefont {S.}~\bibnamefont {Nasri}}, \
  and\ \bibinfo {author} {\bibfnamefont {R.}~\bibnamefont {Soualah}},\ }\href
  {\doibase 10.1103/PhysRevD.89.095010} {\bibfield  {journal} {\bibinfo
  {journal} {Phys. Rev. D}\ }\textbf {\bibinfo {volume} {89}},\ \bibinfo
  {pages} {095010} (\bibinfo {year} {2014})},\ \Eprint
  {http://arxiv.org/abs/1403.5694} {arXiv:1403.5694 [hep-ph]} \BibitemShut
  {NoStop}%
\bibitem [{\citenamefont {Chen}\ \emph {et~al.}(2014)\citenamefont {Chen},
  \citenamefont {McDonald},\ and\ \citenamefont {Nasri}}]{Chen:2014ska}%
  \BibitemOpen
  \bibfield  {author} {\bibinfo {author} {\bibfnamefont {C.-S.}\ \bibnamefont
  {Chen}}, \bibinfo {author} {\bibfnamefont {K.~L.}\ \bibnamefont {McDonald}},
  \ and\ \bibinfo {author} {\bibfnamefont {S.}~\bibnamefont {Nasri}},\ }\href
  {\doibase 10.1016/j.physletb.2014.05.082} {\bibfield  {journal} {\bibinfo
  {journal} {Phys. Lett. B}\ }\textbf {\bibinfo {volume} {734}},\ \bibinfo
  {pages} {388} (\bibinfo {year} {2014})},\ \Eprint
  {http://arxiv.org/abs/1404.6033} {arXiv:1404.6033 [hep-ph]} \BibitemShut
  {NoStop}%
\bibitem [{\citenamefont {Ahriche}\ \emph {et~al.}(2016)\citenamefont
  {Ahriche}, \citenamefont {McDonald},\ and\ \citenamefont
  {Nasri}}]{Ahriche:2015loa}%
  \BibitemOpen
  \bibfield  {author} {\bibinfo {author} {\bibfnamefont {A.}~\bibnamefont
  {Ahriche}}, \bibinfo {author} {\bibfnamefont {K.~L.}\ \bibnamefont
  {McDonald}}, \ and\ \bibinfo {author} {\bibfnamefont {S.}~\bibnamefont
  {Nasri}},\ }\href {\doibase 10.1007/JHEP02(2016)038} {\bibfield  {journal}
  {\bibinfo  {journal} {JHEP}\ }\textbf {\bibinfo {volume} {02}},\ \bibinfo
  {pages} {038} (\bibinfo {year} {2016})},\ \Eprint
  {http://arxiv.org/abs/1508.02607} {arXiv:1508.02607 [hep-ph]} \BibitemShut
  {NoStop}%
\bibitem [{\citenamefont {Herrero-Garcia}\ \emph {et~al.}(2014)\citenamefont
  {Herrero-Garcia}, \citenamefont {Nebot}, \citenamefont {Rius},\ and\
  \citenamefont {Santamaria}}]{Herrero-Garcia:2014hfa}%
  \BibitemOpen
  \bibfield  {author} {\bibinfo {author} {\bibfnamefont {J.}~\bibnamefont
  {Herrero-Garcia}}, \bibinfo {author} {\bibfnamefont {M.}~\bibnamefont
  {Nebot}}, \bibinfo {author} {\bibfnamefont {N.}~\bibnamefont {Rius}}, \ and\
  \bibinfo {author} {\bibfnamefont {A.}~\bibnamefont {Santamaria}},\ }\href
  {\doibase 10.1016/j.nuclphysb.2014.06.001} {\bibfield  {journal} {\bibinfo
  {journal} {Nucl. Phys. B}\ }\textbf {\bibinfo {volume} {885}},\ \bibinfo
  {pages} {542} (\bibinfo {year} {2014})},\ \Eprint
  {http://arxiv.org/abs/1402.4491} {arXiv:1402.4491 [hep-ph]} \BibitemShut
  {NoStop}%
\bibitem [{\citenamefont {Herrero-Garc\'\i{}a}\ \emph
  {et~al.}(2017)\citenamefont {Herrero-Garc\'\i{}a}, \citenamefont {Ohlsson},
  \citenamefont {Riad},\ and\ \citenamefont
  {Wir\'en}}]{Herrero-Garcia:2017xdu}%
  \BibitemOpen
  \bibfield  {author} {\bibinfo {author} {\bibfnamefont {J.}~\bibnamefont
  {Herrero-Garc\'\i{}a}}, \bibinfo {author} {\bibfnamefont {T.}~\bibnamefont
  {Ohlsson}}, \bibinfo {author} {\bibfnamefont {S.}~\bibnamefont {Riad}}, \
  and\ \bibinfo {author} {\bibfnamefont {J.}~\bibnamefont {Wir\'en}},\ }\href
  {\doibase 10.1007/JHEP04(2017)130} {\bibfield  {journal} {\bibinfo  {journal}
  {JHEP}\ }\textbf {\bibinfo {volume} {04}},\ \bibinfo {pages} {130} (\bibinfo
  {year} {2017})},\ \Eprint {http://arxiv.org/abs/1701.05345} {arXiv:1701.05345
  [hep-ph]} \BibitemShut {NoStop}%
\bibitem [{\citenamefont {Centelles~Chuli\'a}\ \emph
  {et~al.}(2018)\citenamefont {Centelles~Chuli\'a}, \citenamefont
  {Srivastava},\ and\ \citenamefont {Valle}}]{CentellesChulia:2018gwr}%
  \BibitemOpen
  \bibfield  {author} {\bibinfo {author} {\bibfnamefont {S.}~\bibnamefont
  {Centelles~Chuli\'a}}, \bibinfo {author} {\bibfnamefont {R.}~\bibnamefont
  {Srivastava}}, \ and\ \bibinfo {author} {\bibfnamefont {J.~W.}\ \bibnamefont
  {Valle}},\ }\href {\doibase 10.1016/j.physletb.2018.03.046} {\bibfield
  {journal} {\bibinfo  {journal} {Phys. Lett. B}\ }\textbf {\bibinfo {volume}
  {781}},\ \bibinfo {pages} {122} (\bibinfo {year} {2018})},\ \Eprint
  {http://arxiv.org/abs/1802.05722} {arXiv:1802.05722 [hep-ph]} \BibitemShut
  {NoStop}%
\bibitem [{\citenamefont {Babu}\ \emph {et~al.}(2020)\citenamefont {Babu},
  \citenamefont {Dev}, \citenamefont {Jana},\ and\ \citenamefont
  {Thapa}}]{Babu:2019mfe}%
  \BibitemOpen
  \bibfield  {author} {\bibinfo {author} {\bibfnamefont {K.}~\bibnamefont
  {Babu}}, \bibinfo {author} {\bibfnamefont {P.~B.}\ \bibnamefont {Dev}},
  \bibinfo {author} {\bibfnamefont {S.}~\bibnamefont {Jana}}, \ and\ \bibinfo
  {author} {\bibfnamefont {A.}~\bibnamefont {Thapa}},\ }\href {\doibase
  10.1007/JHEP03(2020)006} {\bibfield  {journal} {\bibinfo  {journal} {JHEP}\
  }\textbf {\bibinfo {volume} {03}},\ \bibinfo {pages} {006} (\bibinfo {year}
  {2020})},\ \Eprint {http://arxiv.org/abs/1907.09498} {arXiv:1907.09498
  [hep-ph]} \BibitemShut {NoStop}%
\bibitem [{\citenamefont {Nieves}\ and\ \citenamefont
  {Pal}(2004)}]{Nieves:2003in}%
  \BibitemOpen
  \bibfield  {author} {\bibinfo {author} {\bibfnamefont {J.~F.}\ \bibnamefont
  {Nieves}}\ and\ \bibinfo {author} {\bibfnamefont {P.~B.}\ \bibnamefont
  {Pal}},\ }\href {\doibase 10.1119/1.1757445} {\bibfield  {journal} {\bibinfo
  {journal} {Am. J. Phys.}\ }\textbf {\bibinfo {volume} {72}},\ \bibinfo
  {pages} {1100} (\bibinfo {year} {2004})},\ \Eprint
  {http://arxiv.org/abs/hep-ph/0306087} {arXiv:hep-ph/0306087} \BibitemShut
  {NoStop}%
\bibitem [{\citenamefont {Czarnecki}\ \emph {et~al.}(2019)\citenamefont
  {Czarnecki}, \citenamefont {Marciano},\ and\ \citenamefont
  {Sirlin}}]{Czarnecki:2019mwq}%
  \BibitemOpen
  \bibfield  {author} {\bibinfo {author} {\bibfnamefont {A.}~\bibnamefont
  {Czarnecki}}, \bibinfo {author} {\bibfnamefont {W.~J.}\ \bibnamefont
  {Marciano}}, \ and\ \bibinfo {author} {\bibfnamefont {A.}~\bibnamefont
  {Sirlin}},\ }\href {\doibase 10.1103/PhysRevD.100.073008} {\bibfield
  {journal} {\bibinfo  {journal} {Phys. Rev. D}\ }\textbf {\bibinfo {volume}
  {100}},\ \bibinfo {pages} {073008} (\bibinfo {year} {2019})},\ \Eprint
  {http://arxiv.org/abs/1907.06737} {arXiv:1907.06737 [hep-ph]} \BibitemShut
  {NoStop}%
\bibitem [{\citenamefont {Seng}\ \emph {et~al.}(2019)\citenamefont {Seng},
  \citenamefont {Gorchtein},\ and\ \citenamefont
  {Ramsey-Musolf}}]{Seng:2018qru}%
  \BibitemOpen
  \bibfield  {author} {\bibinfo {author} {\bibfnamefont {C.~Y.}\ \bibnamefont
  {Seng}}, \bibinfo {author} {\bibfnamefont {M.}~\bibnamefont {Gorchtein}}, \
  and\ \bibinfo {author} {\bibfnamefont {M.~J.}\ \bibnamefont
  {Ramsey-Musolf}},\ }\href {\doibase 10.1103/PhysRevD.100.013001} {\bibfield
  {journal} {\bibinfo  {journal} {Phys. Rev. D}\ }\textbf {\bibinfo {volume}
  {100}},\ \bibinfo {pages} {013001} (\bibinfo {year} {2019})},\ \Eprint
  {http://arxiv.org/abs/1812.03352} {arXiv:1812.03352 [nucl-th]} \BibitemShut
  {NoStop}%
\bibitem [{\citenamefont {Gorchtein}(2019)}]{Gorchtein:2018fxl}%
  \BibitemOpen
  \bibfield  {author} {\bibinfo {author} {\bibfnamefont {M.}~\bibnamefont
  {Gorchtein}},\ }\href {\doibase 10.1103/PhysRevLett.123.042503} {\bibfield
  {journal} {\bibinfo  {journal} {Phys. Rev. Lett.}\ }\textbf {\bibinfo
  {volume} {123}},\ \bibinfo {pages} {042503} (\bibinfo {year} {2019})},\
  \Eprint {http://arxiv.org/abs/1812.04229} {arXiv:1812.04229 [nucl-th]}
  \BibitemShut {NoStop}%
\bibitem [{\citenamefont {Aoki}\ \emph {et~al.}(2020)\citenamefont {Aoki} \emph
  {et~al.}}]{Aoki:2019cca}%
  \BibitemOpen
  \bibfield  {author} {\bibinfo {author} {\bibfnamefont {S.}~\bibnamefont
  {Aoki}} \emph {et~al.} (\bibinfo {collaboration} {Flavour Lattice Averaging
  Group}),\ }\href {\doibase 10.1140/epjc/s10052-019-7354-7} {\bibfield
  {journal} {\bibinfo  {journal} {Eur. Phys. J. C}\ }\textbf {\bibinfo {volume}
  {80}},\ \bibinfo {pages} {113} (\bibinfo {year} {2020})},\ \Eprint
  {http://arxiv.org/abs/1902.08191} {arXiv:1902.08191 [hep-lat]} \BibitemShut
  {NoStop}%
\bibitem [{\citenamefont {Crivellin}\ \emph
  {et~al.}(2020{\natexlab{c}})\citenamefont {Crivellin}, \citenamefont
  {Hoferichter}, \citenamefont {Manzari},\ and\ \citenamefont
  {Montull}}]{Crivellin:2020zul}%
  \BibitemOpen
  \bibfield  {author} {\bibinfo {author} {\bibfnamefont {A.}~\bibnamefont
  {Crivellin}}, \bibinfo {author} {\bibfnamefont {M.}~\bibnamefont
  {Hoferichter}}, \bibinfo {author} {\bibfnamefont {C.~A.}\ \bibnamefont
  {Manzari}}, \ and\ \bibinfo {author} {\bibfnamefont {M.}~\bibnamefont
  {Montull}},\ }\href {\doibase 10.1103/PhysRevLett.125.091801} {\bibfield
  {journal} {\bibinfo  {journal} {Phys. Rev. Lett.}\ }\textbf {\bibinfo
  {volume} {125}},\ \bibinfo {pages} {091801} (\bibinfo {year}
  {2020}{\natexlab{c}})},\ \Eprint {http://arxiv.org/abs/2003.04886}
  {arXiv:2003.04886 [hep-ph]} \BibitemShut {NoStop}%
\bibitem [{\citenamefont {De~Blas}\ \emph {et~al.}(2020)\citenamefont {De~Blas}
  \emph {et~al.}}]{deBlas:2019okz}%
  \BibitemOpen
  \bibfield  {author} {\bibinfo {author} {\bibfnamefont {J.}~\bibnamefont
  {De~Blas}} \emph {et~al.},\ }\href {\doibase 10.1140/epjc/s10052-020-7904-z}
  {\bibfield  {journal} {\bibinfo  {journal} {Eur. Phys. J. C}\ }\textbf
  {\bibinfo {volume} {80}},\ \bibinfo {pages} {456} (\bibinfo {year} {2020})},\
  \Eprint {http://arxiv.org/abs/1910.14012} {arXiv:1910.14012 [hep-ph]}
  \BibitemShut {NoStop}%
\bibitem [{\citenamefont {Crivellin}\ \emph {et~al.}(2018)\citenamefont
  {Crivellin}, \citenamefont {Hoferichter},\ and\ \citenamefont
  {Schmidt-Wellenburg}}]{Crivellin:2018qmi}%
  \BibitemOpen
  \bibfield  {author} {\bibinfo {author} {\bibfnamefont {A.}~\bibnamefont
  {Crivellin}}, \bibinfo {author} {\bibfnamefont {M.}~\bibnamefont
  {Hoferichter}}, \ and\ \bibinfo {author} {\bibfnamefont {P.}~\bibnamefont
  {Schmidt-Wellenburg}},\ }\href {\doibase 10.1103/PhysRevD.98.113002}
  {\bibfield  {journal} {\bibinfo  {journal} {Phys. Rev. D}\ }\textbf {\bibinfo
  {volume} {98}},\ \bibinfo {pages} {113002} (\bibinfo {year} {2018})},\
  \Eprint {http://arxiv.org/abs/1807.11484} {arXiv:1807.11484 [hep-ph]}
  \BibitemShut {NoStop}%
\bibitem [{\citenamefont {Bertl}\ \emph {et~al.}(2006)\citenamefont {Bertl}
  \emph {et~al.}}]{Bertl:2006up}%
  \BibitemOpen
  \bibfield  {author} {\bibinfo {author} {\bibfnamefont {W.~H.}\ \bibnamefont
  {Bertl}} \emph {et~al.} (\bibinfo {collaboration} {SINDRUM II}),\ }\href
  {\doibase 10.1140/epjc/s2006-02582-x} {\bibfield  {journal} {\bibinfo
  {journal} {Eur. Phys. J. C}\ }\textbf {\bibinfo {volume} {47}},\ \bibinfo
  {pages} {337} (\bibinfo {year} {2006})}\BibitemShut {NoStop}%
\bibitem [{\citenamefont {Aubert}\ \emph {et~al.}(2010)\citenamefont {Aubert}
  \emph {et~al.}}]{Aubert:2009ag}%
  \BibitemOpen
  \bibfield  {author} {\bibinfo {author} {\bibfnamefont {B.}~\bibnamefont
  {Aubert}} \emph {et~al.} (\bibinfo {collaboration} {BaBar}),\ }\href
  {\doibase 10.1103/PhysRevLett.104.021802} {\bibfield  {journal} {\bibinfo
  {journal} {Phys. Rev. Lett.}\ }\textbf {\bibinfo {volume} {104}},\ \bibinfo
  {pages} {021802} (\bibinfo {year} {2010})},\ \Eprint
  {http://arxiv.org/abs/0908.2381} {arXiv:0908.2381 [hep-ex]} \BibitemShut
  {NoStop}%
\bibitem [{\citenamefont {Baldini}\ \emph {et~al.}(2016)\citenamefont {Baldini}
  \emph {et~al.}}]{TheMEG:2016wtm}%
  \BibitemOpen
  \bibfield  {author} {\bibinfo {author} {\bibfnamefont {A.}~\bibnamefont
  {Baldini}} \emph {et~al.} (\bibinfo {collaboration} {MEG}),\ }\href {\doibase
  10.1140/epjc/s10052-016-4271-x} {\bibfield  {journal} {\bibinfo  {journal}
  {Eur. Phys. J. C}\ }\textbf {\bibinfo {volume} {76}},\ \bibinfo {pages} {434}
  (\bibinfo {year} {2016})},\ \Eprint {http://arxiv.org/abs/1605.05081}
  {arXiv:1605.05081 [hep-ex]} \BibitemShut {NoStop}%
\bibitem [{\citenamefont {Andreev}\ \emph {et~al.}(2018)\citenamefont {Andreev}
  \emph {et~al.}}]{Andreev:2018ayy}%
  \BibitemOpen
  \bibfield  {author} {\bibinfo {author} {\bibfnamefont {V.}~\bibnamefont
  {Andreev}} \emph {et~al.} (\bibinfo {collaboration} {ACME}),\ }\href
  {\doibase 10.1038/s41586-018-0599-8} {\bibfield  {journal} {\bibinfo
  {journal} {Nature}\ }\textbf {\bibinfo {volume} {562}},\ \bibinfo {pages}
  {355} (\bibinfo {year} {2018})}\BibitemShut {NoStop}%
\bibitem [{\citenamefont {Bellgardt}\ \emph {et~al.}(1988)\citenamefont
  {Bellgardt} \emph {et~al.}}]{Bellgardt:1987du}%
  \BibitemOpen
  \bibfield  {author} {\bibinfo {author} {\bibfnamefont {U.}~\bibnamefont
  {Bellgardt}} \emph {et~al.} (\bibinfo {collaboration} {SINDRUM}),\ }\href
  {\doibase 10.1016/0550-3213(88)90462-2} {\bibfield  {journal} {\bibinfo
  {journal} {Nucl. Phys. B}\ }\textbf {\bibinfo {volume} {299}},\ \bibinfo
  {pages} {1} (\bibinfo {year} {1988})}\BibitemShut {NoStop}%
\bibitem [{\citenamefont {Hayasaka}\ \emph {et~al.}(2010)\citenamefont
  {Hayasaka} \emph {et~al.}}]{Hayasaka:2010np}%
  \BibitemOpen
  \bibfield  {author} {\bibinfo {author} {\bibfnamefont {K.}~\bibnamefont
  {Hayasaka}} \emph {et~al.},\ }\href {\doibase 10.1016/j.physletb.2010.03.037}
  {\bibfield  {journal} {\bibinfo  {journal} {Phys. Lett. B}\ }\textbf
  {\bibinfo {volume} {687}},\ \bibinfo {pages} {139} (\bibinfo {year}
  {2010})},\ \Eprint {http://arxiv.org/abs/1001.3221} {arXiv:1001.3221
  [hep-ex]} \BibitemShut {NoStop}%
\bibitem [{\citenamefont {Lees}\ \emph {et~al.}(2010)\citenamefont {Lees} \emph
  {et~al.}}]{Lees:2010ez}%
  \BibitemOpen
  \bibfield  {author} {\bibinfo {author} {\bibfnamefont {J.}~\bibnamefont
  {Lees}} \emph {et~al.} (\bibinfo {collaboration} {BaBar}),\ }\href {\doibase
  10.1103/PhysRevD.81.111101} {\bibfield  {journal} {\bibinfo  {journal} {Phys.
  Rev. D}\ }\textbf {\bibinfo {volume} {81}},\ \bibinfo {pages} {111101}
  (\bibinfo {year} {2010})},\ \Eprint {http://arxiv.org/abs/1002.4550}
  {arXiv:1002.4550 [hep-ex]} \BibitemShut {NoStop}%
\bibitem [{\citenamefont {Aaij}\ \emph
  {et~al.}(2015{\natexlab{c}})\citenamefont {Aaij} \emph
  {et~al.}}]{Aaij:2014azz}%
  \BibitemOpen
  \bibfield  {author} {\bibinfo {author} {\bibfnamefont {R.}~\bibnamefont
  {Aaij}} \emph {et~al.} (\bibinfo {collaboration} {LHCb}),\ }\href {\doibase
  10.1007/JHEP02(2015)121} {\bibfield  {journal} {\bibinfo  {journal} {JHEP}\
  }\textbf {\bibinfo {volume} {02}},\ \bibinfo {pages} {121} (\bibinfo {year}
  {2015}{\natexlab{c}})},\ \Eprint {http://arxiv.org/abs/1409.8548}
  {arXiv:1409.8548 [hep-ex]} \BibitemShut {NoStop}%
\bibitem [{\citenamefont {Cao}\ \emph {et~al.}(2018)\citenamefont {Cao},
  \citenamefont {Li}, \citenamefont {Xie},\ and\ \citenamefont
  {Zhang}}]{Cao:2017ffm}%
  \BibitemOpen
  \bibfield  {author} {\bibinfo {author} {\bibfnamefont {Q.-H.}\ \bibnamefont
  {Cao}}, \bibinfo {author} {\bibfnamefont {G.}~\bibnamefont {Li}}, \bibinfo
  {author} {\bibfnamefont {K.-P.}\ \bibnamefont {Xie}}, \ and\ \bibinfo
  {author} {\bibfnamefont {J.}~\bibnamefont {Zhang}},\ }\href {\doibase
  10.1103/PhysRevD.97.115036} {\bibfield  {journal} {\bibinfo  {journal} {Phys.
  Rev. D}\ }\textbf {\bibinfo {volume} {97}},\ \bibinfo {pages} {115036}
  (\bibinfo {year} {2018})},\ \Eprint {http://arxiv.org/abs/1711.02113}
  {arXiv:1711.02113 [hep-ph]} \BibitemShut {NoStop}%
\bibitem [{\citenamefont {Alcaide}\ \emph {et~al.}(2018)\citenamefont
  {Alcaide}, \citenamefont {Chala},\ and\ \citenamefont
  {Santamaria}}]{Alcaide:2017dcx}%
  \BibitemOpen
  \bibfield  {author} {\bibinfo {author} {\bibfnamefont {J.}~\bibnamefont
  {Alcaide}}, \bibinfo {author} {\bibfnamefont {M.}~\bibnamefont {Chala}}, \
  and\ \bibinfo {author} {\bibfnamefont {A.}~\bibnamefont {Santamaria}},\
  }\href {\doibase 10.1016/j.physletb.2018.02.001} {\bibfield  {journal}
  {\bibinfo  {journal} {Phys. Lett. B}\ }\textbf {\bibinfo {volume} {779}},\
  \bibinfo {pages} {107} (\bibinfo {year} {2018})},\ \Eprint
  {http://arxiv.org/abs/1710.05885} {arXiv:1710.05885 [hep-ph]} \BibitemShut
  {NoStop}%
\bibitem [{\citenamefont {Alcaide}\ and\ \citenamefont
  {Mileo}(2020)}]{Alcaide:2019kdr}%
  \BibitemOpen
  \bibfield  {author} {\bibinfo {author} {\bibfnamefont {J.}~\bibnamefont
  {Alcaide}}\ and\ \bibinfo {author} {\bibfnamefont {N.~I.}\ \bibnamefont
  {Mileo}},\ }\href {\doibase 10.1103/PhysRevD.102.075030} {\bibfield
  {journal} {\bibinfo  {journal} {Phys. Rev. D}\ }\textbf {\bibinfo {volume}
  {102}},\ \bibinfo {pages} {075030} (\bibinfo {year} {2020})},\ \Eprint
  {http://arxiv.org/abs/1906.08685} {arXiv:1906.08685 [hep-ph]} \BibitemShut
  {NoStop}%
\bibitem [{\citenamefont {Aad}\ \emph {et~al.}(2020)\citenamefont {Aad} \emph
  {et~al.}}]{Aad:2019vnb}%
  \BibitemOpen
  \bibfield  {author} {\bibinfo {author} {\bibfnamefont {G.}~\bibnamefont
  {Aad}} \emph {et~al.} (\bibinfo {collaboration} {ATLAS}),\ }\href {\doibase
  10.1140/epjc/s10052-019-7594-6} {\bibfield  {journal} {\bibinfo  {journal}
  {Eur. Phys. J. C}\ }\textbf {\bibinfo {volume} {80}},\ \bibinfo {pages} {123}
  (\bibinfo {year} {2020})},\ \Eprint {http://arxiv.org/abs/1908.08215}
  {arXiv:1908.08215 [hep-ex]} \BibitemShut {NoStop}%
\bibitem [{\citenamefont {Alwall}\ \emph {et~al.}(2014)\citenamefont {Alwall},
  \citenamefont {Frederix}, \citenamefont {Frixione}, \citenamefont {Hirschi},
  \citenamefont {Maltoni}, \citenamefont {Mattelaer}, \citenamefont {Shao},
  \citenamefont {Stelzer}, \citenamefont {Torrielli},\ and\ \citenamefont
  {Zaro}}]{Alwall:2014hca}%
  \BibitemOpen
  \bibfield  {author} {\bibinfo {author} {\bibfnamefont {J.}~\bibnamefont
  {Alwall}}, \bibinfo {author} {\bibfnamefont {R.}~\bibnamefont {Frederix}},
  \bibinfo {author} {\bibfnamefont {S.}~\bibnamefont {Frixione}}, \bibinfo
  {author} {\bibfnamefont {V.}~\bibnamefont {Hirschi}}, \bibinfo {author}
  {\bibfnamefont {F.}~\bibnamefont {Maltoni}}, \bibinfo {author} {\bibfnamefont
  {O.}~\bibnamefont {Mattelaer}}, \bibinfo {author} {\bibfnamefont {H.~S.}\
  \bibnamefont {Shao}}, \bibinfo {author} {\bibfnamefont {T.}~\bibnamefont
  {Stelzer}}, \bibinfo {author} {\bibfnamefont {P.}~\bibnamefont {Torrielli}},
  \ and\ \bibinfo {author} {\bibfnamefont {M.}~\bibnamefont {Zaro}},\ }\href
  {\doibase 10.1007/JHEP07(2014)079} {\bibfield  {journal} {\bibinfo  {journal}
  {JHEP}\ }\textbf {\bibinfo {volume} {07}},\ \bibinfo {pages} {079} (\bibinfo
  {year} {2014})},\ \Eprint {http://arxiv.org/abs/1405.0301} {arXiv:1405.0301
  [hep-ph]} \BibitemShut {NoStop}%
\bibitem [{\citenamefont {Abdallah}\ \emph {et~al.}(2005)\citenamefont
  {Abdallah} \emph {et~al.}}]{Abdallah:2003np}%
  \BibitemOpen
  \bibfield  {author} {\bibinfo {author} {\bibfnamefont {J.}~\bibnamefont
  {Abdallah}} \emph {et~al.} (\bibinfo {collaboration} {DELPHI}),\ }\href
  {\doibase 10.1140/epjc/s2004-02051-8} {\bibfield  {journal} {\bibinfo
  {journal} {Eur. Phys. J. C}\ }\textbf {\bibinfo {volume} {38}},\ \bibinfo
  {pages} {395} (\bibinfo {year} {2005})},\ \Eprint
  {http://arxiv.org/abs/hep-ex/0406019} {arXiv:hep-ex/0406019} \BibitemShut
  {NoStop}%
\bibitem [{\citenamefont {Abdallah}\ \emph {et~al.}(2009)\citenamefont
  {Abdallah} \emph {et~al.}}]{Abdallah:2008aa}%
  \BibitemOpen
  \bibfield  {author} {\bibinfo {author} {\bibfnamefont {J.}~\bibnamefont
  {Abdallah}} \emph {et~al.} (\bibinfo {collaboration} {DELPHI}),\ }\href
  {\doibase 10.1140/epjc/s10052-009-0874-9} {\bibfield  {journal} {\bibinfo
  {journal} {Eur. Phys. J. C}\ }\textbf {\bibinfo {volume} {60}},\ \bibinfo
  {pages} {17} (\bibinfo {year} {2009})},\ \Eprint
  {http://arxiv.org/abs/0901.4486} {arXiv:0901.4486 [hep-ex]} \BibitemShut
  {NoStop}%
\bibitem [{\citenamefont {Fox}\ \emph {et~al.}(2011)\citenamefont {Fox},
  \citenamefont {Harnik}, \citenamefont {Kopp},\ and\ \citenamefont
  {Tsai}}]{Fox:2011fx}%
  \BibitemOpen
  \bibfield  {author} {\bibinfo {author} {\bibfnamefont {P.~J.}\ \bibnamefont
  {Fox}}, \bibinfo {author} {\bibfnamefont {R.}~\bibnamefont {Harnik}},
  \bibinfo {author} {\bibfnamefont {J.}~\bibnamefont {Kopp}}, \ and\ \bibinfo
  {author} {\bibfnamefont {Y.}~\bibnamefont {Tsai}},\ }\href {\doibase
  10.1103/PhysRevD.84.014028} {\bibfield  {journal} {\bibinfo  {journal} {Phys.
  Rev. D}\ }\textbf {\bibinfo {volume} {84}},\ \bibinfo {pages} {014028}
  (\bibinfo {year} {2011})},\ \Eprint {http://arxiv.org/abs/1103.0240}
  {arXiv:1103.0240 [hep-ph]} \BibitemShut {NoStop}%
\bibitem [{Apo(2017)}]{ApollinariG.:2017ojx}%
  \BibitemOpen
  \href {\doibase 10.23731/CYRM-2017-004} {\ \textbf {\bibinfo {volume}
  {4/2017}} (\bibinfo {year} {2017}),\ 10.23731/CYRM-2017-004}\BibitemShut
  {NoStop}%
\bibitem [{\citenamefont {Abada}\ \emph
  {et~al.}(2019{\natexlab{a}})\citenamefont {Abada} \emph
  {et~al.}}]{Benedikt:2018csr}%
  \BibitemOpen
  \bibfield  {author} {\bibinfo {author} {\bibfnamefont {A.}~\bibnamefont
  {Abada}} \emph {et~al.} (\bibinfo {collaboration} {FCC}),\ }\href {\doibase
  10.1140/epjst/e2019-900087-0} {\bibfield  {journal} {\bibinfo  {journal}
  {Eur. Phys. J. ST}\ }\textbf {\bibinfo {volume} {228}},\ \bibinfo {pages}
  {755} (\bibinfo {year} {2019}{\natexlab{a}})}\BibitemShut {NoStop}%
\bibitem [{\citenamefont {Baumholzer}\ \emph {et~al.}(2020)\citenamefont
  {Baumholzer}, \citenamefont {Brdar}, \citenamefont {Schwaller},\ and\
  \citenamefont {Segner}}]{Baumholzer:2019twf}%
  \BibitemOpen
  \bibfield  {author} {\bibinfo {author} {\bibfnamefont {S.}~\bibnamefont
  {Baumholzer}}, \bibinfo {author} {\bibfnamefont {V.}~\bibnamefont {Brdar}},
  \bibinfo {author} {\bibfnamefont {P.}~\bibnamefont {Schwaller}}, \ and\
  \bibinfo {author} {\bibfnamefont {A.}~\bibnamefont {Segner}},\ }\href
  {\doibase 10.1007/JHEP09(2020)136} {\bibfield  {journal} {\bibinfo  {journal}
  {JHEP}\ }\textbf {\bibinfo {volume} {09}},\ \bibinfo {pages} {136} (\bibinfo
  {year} {2020})},\ \Eprint {http://arxiv.org/abs/1912.08215} {arXiv:1912.08215
  [hep-ph]} \BibitemShut {NoStop}%
\bibitem [{\citenamefont {Inami}(2016)}]{Inami:2016aba}%
  \BibitemOpen
  \bibfield  {author} {\bibinfo {author} {\bibfnamefont {K.}~\bibnamefont
  {Inami}} (\bibinfo {collaboration} {Belle-II}),\ }\href {\doibase
  10.22323/1.282.0574} {\bibfield  {journal} {\bibinfo  {journal} {PoS}\
  }\textbf {\bibinfo {volume} {ICHEP2016}},\ \bibinfo {pages} {574} (\bibinfo
  {year} {2016})}\BibitemShut {NoStop}%
\bibitem [{\citenamefont {Pich}(2020)}]{Pich:2020qna}%
  \BibitemOpen
  \bibfield  {author} {\bibinfo {author} {\bibfnamefont {A.}~\bibnamefont
  {Pich}},\ }\href@noop {} {\  (\bibinfo {year} {2020})},\ \Eprint
  {http://arxiv.org/abs/2012.07099} {arXiv:2012.07099 [hep-ph]} \BibitemShut
  {NoStop}%
\bibitem [{\citenamefont {Baer}\ \emph {et~al.}(2013)\citenamefont {Baer} \emph
  {et~al.}}]{Baer:2013cma}%
  \BibitemOpen
  \bibfield  {author} {\bibinfo {author} {\bibfnamefont {H.}~\bibnamefont
  {Baer}} \emph {et~al.},\ }\href@noop {} {\  (\bibinfo {year} {2013})},\
  \Eprint {http://arxiv.org/abs/1306.6352} {arXiv:1306.6352 [hep-ph]}
  \BibitemShut {NoStop}%
\bibitem [{\citenamefont {Aicheler}\ \emph {et~al.}(2012)\citenamefont
  {Aicheler} \emph {et~al.}}]{Aicheler:2012bya}%
  \BibitemOpen
  \bibfield  {author} {\bibinfo {author} {\bibfnamefont {M.}~\bibnamefont
  {Aicheler}} \emph {et~al.},\ }\href {\doibase 10.5170/CERN-2012-007} {\
  (\bibinfo {year} {2012}),\ 10.5170/CERN-2012-007}\BibitemShut {NoStop}%
\bibitem [{\citenamefont {An}\ \emph {et~al.}(2019)\citenamefont {An} \emph
  {et~al.}}]{An:2018dwb}%
  \BibitemOpen
  \bibfield  {author} {\bibinfo {author} {\bibfnamefont {F.}~\bibnamefont {An}}
  \emph {et~al.},\ }\href {\doibase 10.1088/1674-1137/43/4/043002} {\bibfield
  {journal} {\bibinfo  {journal} {Chin. Phys. C}\ }\textbf {\bibinfo {volume}
  {43}},\ \bibinfo {pages} {043002} (\bibinfo {year} {2019})},\ \Eprint
  {http://arxiv.org/abs/1810.09037} {arXiv:1810.09037 [hep-ex]} \BibitemShut
  {NoStop}%
\bibitem [{\citenamefont {Abada}\ \emph
  {et~al.}(2019{\natexlab{b}})\citenamefont {Abada} \emph
  {et~al.}}]{Abada:2019zxq}%
  \BibitemOpen
  \bibfield  {author} {\bibinfo {author} {\bibfnamefont {A.}~\bibnamefont
  {Abada}} \emph {et~al.} (\bibinfo {collaboration} {FCC}),\ }\href {\doibase
  10.1140/epjst/e2019-900045-4} {\bibfield  {journal} {\bibinfo  {journal}
  {Eur. Phys. J. ST}\ }\textbf {\bibinfo {volume} {228}},\ \bibinfo {pages}
  {261} (\bibinfo {year} {2019}{\natexlab{b}})}\BibitemShut {NoStop}%
\bibitem [{\citenamefont {Habermehl}\ \emph {et~al.}(2020)\citenamefont
  {Habermehl}, \citenamefont {Berggren},\ and\ \citenamefont
  {List}}]{Habermehl:2020njb}%
  \BibitemOpen
  \bibfield  {author} {\bibinfo {author} {\bibfnamefont {M.}~\bibnamefont
  {Habermehl}}, \bibinfo {author} {\bibfnamefont {M.}~\bibnamefont {Berggren}},
  \ and\ \bibinfo {author} {\bibfnamefont {J.}~\bibnamefont {List}},\ }\href
  {\doibase 10.1103/PhysRevD.101.075053} {\bibfield  {journal} {\bibinfo
  {journal} {Phys. Rev. D}\ }\textbf {\bibinfo {volume} {101}},\ \bibinfo
  {pages} {075053} (\bibinfo {year} {2020})},\ \Eprint
  {http://arxiv.org/abs/2001.03011} {arXiv:2001.03011 [hep-ex]} \BibitemShut
  {NoStop}%
\bibitem [{\citenamefont {Liu}\ \emph {et~al.}(2019)\citenamefont {Liu},
  \citenamefont {Xu},\ and\ \citenamefont {Zhang}}]{Liu:2019ogn}%
  \BibitemOpen
  \bibfield  {author} {\bibinfo {author} {\bibfnamefont {Z.}~\bibnamefont
  {Liu}}, \bibinfo {author} {\bibfnamefont {Y.-H.}\ \bibnamefont {Xu}}, \ and\
  \bibinfo {author} {\bibfnamefont {Y.}~\bibnamefont {Zhang}},\ }\href
  {\doibase 10.1007/JHEP06(2019)009} {\bibfield  {journal} {\bibinfo  {journal}
  {JHEP}\ }\textbf {\bibinfo {volume} {06}},\ \bibinfo {pages} {009} (\bibinfo
  {year} {2019})},\ \Eprint {http://arxiv.org/abs/1903.12114} {arXiv:1903.12114
  [hep-ph]} \BibitemShut {NoStop}%
\bibitem [{\citenamefont {Kitano}\ \emph {et~al.}(2002)\citenamefont {Kitano},
  \citenamefont {Koike},\ and\ \citenamefont {Okada}}]{Kitano:2002mt}%
  \BibitemOpen
  \bibfield  {author} {\bibinfo {author} {\bibfnamefont {R.}~\bibnamefont
  {Kitano}}, \bibinfo {author} {\bibfnamefont {M.}~\bibnamefont {Koike}}, \
  and\ \bibinfo {author} {\bibfnamefont {Y.}~\bibnamefont {Okada}},\ }\href
  {\doibase 10.1103/PhysRevD.76.059902} {\bibfield  {journal} {\bibinfo
  {journal} {Phys. Rev. D}\ }\textbf {\bibinfo {volume} {66}},\ \bibinfo
  {pages} {096002} (\bibinfo {year} {2002})},\ \bibinfo {note} {[Erratum:
  Phys.Rev.D 76, 059902 (2007)]},\ \Eprint
  {http://arxiv.org/abs/hep-ph/0203110} {arXiv:hep-ph/0203110} \BibitemShut
  {NoStop}%
\bibitem [{\citenamefont {Suzuki}\ \emph {et~al.}(1987)\citenamefont {Suzuki},
  \citenamefont {Measday},\ and\ \citenamefont {Roalsvig}}]{Suzuki:1987jf}%
  \BibitemOpen
  \bibfield  {author} {\bibinfo {author} {\bibfnamefont {T.}~\bibnamefont
  {Suzuki}}, \bibinfo {author} {\bibfnamefont {D.~F.}\ \bibnamefont {Measday}},
  \ and\ \bibinfo {author} {\bibfnamefont {J.}~\bibnamefont {Roalsvig}},\
  }\href {\doibase 10.1103/PhysRevC.35.2212} {\bibfield  {journal} {\bibinfo
  {journal} {Phys. Rev. C}\ }\textbf {\bibinfo {volume} {35}},\ \bibinfo
  {pages} {2212} (\bibinfo {year} {1987})}\BibitemShut {NoStop}%
\end{thebibliography}%

\end{document}